\documentclass[twocolumn]{aastex63}

\usepackage[utf8]{inputenc}
\usepackage{graphicx}	
\usepackage{amsmath}	
\usepackage{amssymb}	
\usepackage{booktabs}
\usepackage{epstopdf}
\usepackage[caption=false]{subfig}
\usepackage{footnote}
\usepackage{multirow}
\usepackage{threeparttable}

\DeclareUnicodeCharacter{0301}{\'{e}}
\DeclareUnicodeCharacter{0303}{\'{i}}
\received{01.02.2021}
\revised{01.03.2021}
\accepted{05.03.2021}

\submitjournal{ApJ}

\shorttitle{Unfolding the thermal side of ram pressure stripping}
\shortauthors{Campitiello et al.}
\graphicspath{{./}{figures/}}

\begin{document}

\title{GASP XXXIV: Unfolding the thermal side of ram pressure stripping in the jellyfish galaxy JO201}

\correspondingauthor{M. Giulia Campitiello}
\email{maria.campitiello2@unibo.it}

\author[0000-0001-5581-3349]{M. Giulia Campitiello}
\affiliation{Dipartimento di Fisica e Astronomia, Università di Bologna, via Piero Gobetti 93/2, 40129 Bologna, Italy}
\affiliation{INAF, Osservatorio Astronomico di Bologna, via Piero Gobetti, 93/3, 40129 Bologna, Italy}

\author[0000-0003-1581-0092]{Alessandro Ignesti}
\affiliation{INAF-Padova Astronomical Observatory, Vicolo dell’Osservatorio 5, I-35122 Padova, Italy}
\affiliation{Dipartimento di Fisica e Astronomia, Università di Bologna, via Piero Gobetti 93/2, 40129 Bologna, Italy}

\author[0000-0002-0843-3009]{Myriam Gitti}
\affiliation{Dipartimento di Fisica e Astronomia, Università di Bologna, via Piero Gobetti 93/2, 40129 Bologna, Italy}
\affiliation{INAF, Istituto di Radioastronomia di Bologna, via Piero Gobetti 101, 40129 Bologna, Italy}

\author[0000-0001-9807-8479]{Fabrizio Brighenti}
\affiliation{Dipartimento di Fisica e Astronomia, Università di Bologna, via Piero Gobetti 93/2, 40129 Bologna, Italy}
\affiliation{Department of Astronomy and Astrophysics, University of California, 1156 High Street, Santa Cruz, CA 95064, USA}

\author[0000-0002-3585-866X]{Mario Radovich}
\affiliation{INAF-Padova Astronomical Observatory, Vicolo dell’Osservatorio 5, I-35122 Padova, Italy}

\author[0000-0001-5840-9835]{Anna Wolter}
\affiliation{INAF-Osservatorio Astronomico di Brera, via Brera 28, I-20121 Milano, Italy}

\author[0000-0002-8238-9210]{Neven Tomičić}
\affiliation{INAF-Padova Astronomical Observatory, Vicolo dell’Osservatorio 5, I-35122 Padova, Italy}

\author[0000-0002-6179-8007]{Callum Bellhouse}
\affiliation{INAF-Padova Astronomical Observatory, Vicolo dell’Osservatorio 5, I-35122 Padova, Italy}
\affiliation{University of Birmingham School of Physics and Astronomy, Edgbaston, Birmingham B15 2TT, England}

\author[0000-0001-8751-8360]{Bianca M. Poggianti}
\affiliation{INAF-Padova Astronomical Observatory, Vicolo dell’Osservatorio 5, I-35122 Padova, Italy}

\author[0000-0002-1688-482X]{Alessia Moretti}
\affiliation{INAF-Padova Astronomical Observatory, Vicolo dell’Osservatorio 5, I-35122 Padova, Italy}

\author[0000-0003-0980-1499]{Benedetta Vulcani}
\affiliation{INAF-Padova Astronomical Observatory, Vicolo dell’Osservatorio 5, I-35122 Padova, Italy}

\author[0000-0003-2150-1130]{Yara L. Jaff\'e}
\affiliation{Instituto de Física y Astronomía, Universidad de Valparaíso, Avda. Gran Bretaña 1111 Valparaíso, Chile}

\author[0000-0001-9143-6026]{Rosita Paladino}
\affiliation{INAF, Istituto di Radioastronomia di Bologna, via Piero Gobetti 101, 40129 Bologna, Italy}

\author[0000-0001-9184-7845]{Ancla M\"uller}
\affiliation{Ruhr University Bochum, Faculty of Physics and Astronomy, Astronomical Institute, Universit\"atsst 150, 44801 Bochum, Germany}

\author[0000-0002-7042-1965]{Jacopo Fritz}
\affiliation{Instituto de Radioastronomía y Astrofísica, UNAM, Campus Morelia, A.P. 3-72, C.P. 58089, Mexico}

\author[0000-0002-4393-7798]{Ana C. C. Louren\c{c}o}
\affiliation{Instituto de Física y Astronomía, Universidad de Valparaíso, Avda. Gran Bretaña 1111 Valparaíso, Chile}

\author[0000-0002-7296-9780]{Marco Gullieuszik}
\affiliation{INAF-Padova Astronomical Observatory, Vicolo dell’Osservatorio 5, I-35122 Padova, Italy}


\begin{abstract}
X-ray studies of jellyfish galaxies play a crucial role in understanding the interactions between the interstellar medium (ISM) and the intracluster medium (ICM). In this paper, we focused on the jellyfish galaxy JO201. By combining archival Chandra observations, MUSE H$\alpha$ cubes, and maps of the emission fraction of the diffuse ionised gas, we investigated both its high energy spectral properties and the spatial correlation between its X-ray and optical emissions. The X-ray emission of JO201 is provided by both the Compton thick AGN (L$_{\text{X}}^{0.5-10 \text{keV}}$=2.7$\cdot$10$^{41}$ erg s$^{-1}$, not corrected for intrinsic absorption) and an extended component (L$_{\text{X}}^{0.5-10 \, \text{keV}}\approx$ 1.9-4.5$\cdot$10$^{41}$ erg s$^{-1}$) produced by a warm plasma (kT$\approx$1 keV), whose luminosity is higher than expected from the observed star formation (L$_{\text{X}}\sim$3.8$\cdot10^{40}$ erg s$^{-1}$). The spectral analysis showed that the X-ray emission is consistent with the thermal cooling of hot plasma. These properties are similar to the ones found in other jellyfish galaxies showing extended X-ray emission. A point-to-point analysis revealed that this X-ray emission closely follows the ISM distribution, whereas CLOUDY simulations proved that the ionisation triggered by this warm plasma would be able to reproduce the [OI]/H$\alpha$ excess observed in JO201. We conclude that the galactic X-ray emitting plasma is originated on the surface of the ISM as a result of the ICM-ISM interplay. This process would entail the cooling and accretion of the ICM onto the galaxy, which could additionally fuel the star formation, and the emergence of [OI]/H$\alpha$ excess in the optical spectrum. 
\end{abstract}


\section{Introduction}\label{introduzione}
The evolution of galaxies is strongly influenced by all those phenomena that can alter their gas content. Due to internal processes, like feedback from supernovae (SN) and active galactic nuclei (AGN) \citep[e.g.,][]{Ho2014}, or external processes, like harassment \citep[e.g.,][]{Moore1998}, strangulation \citep[e.g.,][]{Larson1980}, and tidal interactions \citep[e.g.,][]{Springel2000}, galaxies can either acquire gas, increasing their star formation (SF), or, loose their gas, turning into passive systems \citep[e.g.,][and references therein]{vanGorkom_2004}. One of the main processes involved in removing gas from galaxies is ram pressure stripping (RPS). This mechanism was proposed for the first time by \cite{Gunn1972} to explain the lack of gas-rich galaxies in clusters. Anytime a galaxy falls into the intracluster medium (ICM), it experiences a force in the opposite direction of its relative motion. If this force overcomes the gravitational one, the gas component is stripped away, causing a transformation that can turn the galaxy into a quenched system. The condition for gas loss is given by $\rho_{\text{ICM}} v^2_{\text{in}}>2 \pi G \Sigma_{\text{s}} \Sigma_{\text{g}}$, that is when the ram pressure is higher than the gravitational pressure. Here  $\Sigma_{\text{s}}$ and $\Sigma_{\text{g}}$ are, respectively, the surface density of stars and gas,  $\rho_{\text{ICM}}$ is the ICM mass density and $v_{\text{in}}$ the in-fall velocity of the galaxy. Starting from this relation, it is possible to identify regions of the phase-space diagram in which the stripping process is heavily favoured; they are characterised by small cluster radii (where the ICM density is higher) and high galaxy velocity \citep{Yoon2017, Jaffe2015,Jaffe2016}. In these regions in particular, the presence of peculiar galaxies was observed: these objects show clear signs of the ongoing stripping process, i. e. tails or "tentacles" of diffuse gas extended in the opposite direction of the galaxy motion \citep{Jaffe2018,Gulli2020}. The most extreme examples of galaxies undergoing strong ram pressure are the so called jellyfish galaxies \citep{Fumagalli2014, Smith2010, Ebeling2014,Poggianti2017}, objects that show extra-planar, unilateral debris visible in the optical/UV light and striking tails of H$\alpha$ ionised gas. \\

Jellyfish galaxies represent the transitional phase between infalling star-forming spirals and quenched cluster early-type galaxies and provide a unique opportunity to understand the impact of gas removal processes on both the star formation and the AGN activity. One of the most recent research projects in this field is GAs Stripping Phenomena (GASP), an European Southern Observatory (ESO) Large Program carried out with the Multi Unit Spectroscopic
Explorer (MUSE) at the Very Large
Telescope (VLT), to observe 94 stripping candidates at z=0.04-0.07 \citep{Poggianti2017}. These galaxies were selected by visual inspection based on the presence of ram-pressure stripping signatures, such as displaced tails and asymmetric morphologies, from the WIdefield Nearby Galaxy-cluster Survey \citep[WINGS][]{Fasano2006,Moretti2014}, its extension OmegaWINGS \citep{Gullieuszik2015,Moretti2017} and the Padova-Millennium Galaxy and Group Catalogue \citep{Calvi2011}.\\

To investigate the effects and the features of such a complex process as the ram pressure stripping, multi-wavelength studies are required. With this approach it is in fact possible, to observe the different phases of the stripped gas \citep[e.g.][]{Poggianti2019}, to reconstruct the star formation history of the galaxy \citep[visible in the optical band, e.g.,][]{Bellhouse2019} and to understand the impact of the magnetic fields on the stripping process \citep[visible in the radio band, e.g.][]{Muller2020}. However, the high-energy side of jellyfish galaxies is still deeply unexplored, with only few objects studied in details: ESO 137-001 and ESO 137-002  in Abell 3627 \citep[][]{Sun2010,Zhang2013}, the GASP galaxy JW100 in Abell 2626 \citep[]{Poggianti2019} and also NGC 6872 in the galaxy group Pavo \citep{Machacek2005}, NGC 4569 in Virgo \citep{Tschoke2001} and UGC 6697 in A1367 \citep{Sun2005}, although for these three latter galaxies a further contribution arising from tidal interactions cannot be completely excluded. These studies found that the galactic extended X-ray emission is produced by a warm plasma with temperature in the range kT $\sim$ 0.7 - 1 keV, likely originated by the interplay between the interstellar medium (ISM) and the ICM on the surface of the stripped tails. From the analysis of JW100, it was observed that the star formation rate (SFR) is not able to explain the observed X-ray luminosity of the galaxy and, thus, an additional emission mechanism is required. Moreover, a spatial correlation between X-ray and H$\alpha$ surface brightness was found and was interpreted as evidence of the strict connection between the X-ray emitting plasma and the ISM. Based on these results and on previous studies about the phase transformation of the
ram pressure stripped gas through shocks, heat conduction, magneto-hydrodynamic
waves and turbulence \citep[e.g.,][]{Cowie1977, Nulsen1982, Gavazzi2001, Sun2006, Sun2007, Sun2010, Fossati2016, Boselli2016}, it was argued that the origin of the extended X-ray emission observed in jellyfish galaxies could be related to the ongoing stripping process, with the X-ray emitting plasma being the results of the complex ICM-ISM interaction triggered by the stripping, which causes either the heating of the ISM through shocks and conduction, or the cooling of the ICM onto the galaxy or the ICM-ISM mixing. \\

With the aim to expand the sample of jellyfish galaxies with detailed X-ray analysis, and with the purpose to address the open questions triggered by the JW100 study, we carried out an X-ray investigation of another jellyfish galaxy of the GASP sample, JO201. This galaxy has been extensively studied at many wavelengths \citep[e.g.,][]{Bellhouse2017, Bellhouse2019, Bellhouse2020,George2018,Moretti2020,Ramatsoku2020} and disposes of deep, archival Chandra observations, necessary for a detailed X-ray analysis. This paper is structured as follows. In Section 2 we present the main properties of JO201 and its host cluster Abell 85; in Section 3 we describe the procedures adopted for the Chandra data reduction and extraction of spectra. In Section 4 we present the results of our detailed spectral analysis, and discuss them in Section 5. Conclusions are presented in Section 6. Throughout this paper, we assume H$_0$= 70 km s$^{-1}$ Mpc$^{-1}$, $\Omega_M=0.3$ and $\Omega_{\Lambda}$= 0.7 (1 arcsec = 1.085 kpc at z = 0.05586) and errors are given at the 1$\sigma$ level.

\section{JO201 and the cluster Abell 85}
As mentioned in the Introduction, to fully understand the ongoing stripping process in JO201, it is necessary to use a multi-wavelength approach and to take into account the properties of the environment in which the galaxy is located. For this reason, we briefly present in the following subsection the results of previous studies of both JO201 and its host cluster Abell 85 (A85). 

\subsection{The galaxy cluster Abell 85}\label{Abell85}
Abell 85 \citep[z=0.05586 and M$_{200}$=1.58 $\cdot 10^{15}$$ M_{\odot}$,][]{Moretti2017} is one of the brightest galaxy clusters in the X-ray sky \citep{Edge1990} and hosts the largest brightest central galaxy (BCG) ever observed in the optical band \citep{Lopez2014}, namely Holm 15A. Despite being classified as a cool-core cluster, A85 is undergoing a merger with two sub-clusters: one from the south (S) and the other one from the south-west (SW). In order to investigate the peculiar dynamical state of A85, \cite{Ichinohe2015} analysed its X-ray emission by  means of a deep Chandra exposure. This analysis highlighted the presence of an apparent brightness excess spiral, starting north of the core and extending counter-clockwise outward from the core out to $\sim$600 kpc. This feature is attributable to the sloshing of the ICM in the gravitational potential of the cluster triggered by previous merger events. Furthermore, it was observed that the S sub-cluster core is almost entirely stripped of the low-entropy gas, providing a case of efficient destruction of a cool core during a merger. Starting from the X-ray spectral analysis, radial profiles of the thermal properties of the cluster were derived. According to them, in the annular region between r$_{\text{1}}$=200 kpc and r$_{\text{2}}$=400 kpc, where JO201 is located, a temperature of kT$\simeq 6.7$ keV, a density of n$_{\text{e}}\simeq$ 10$^{-3}$cm$^{-3}$ and an ICM pressure of p$\simeq$10$^{-2}$ keV cm$^{-3}$ were measured.
 \begin{figure*}
  \hspace{-1.5cm}
   \includegraphics[scale=0.50]{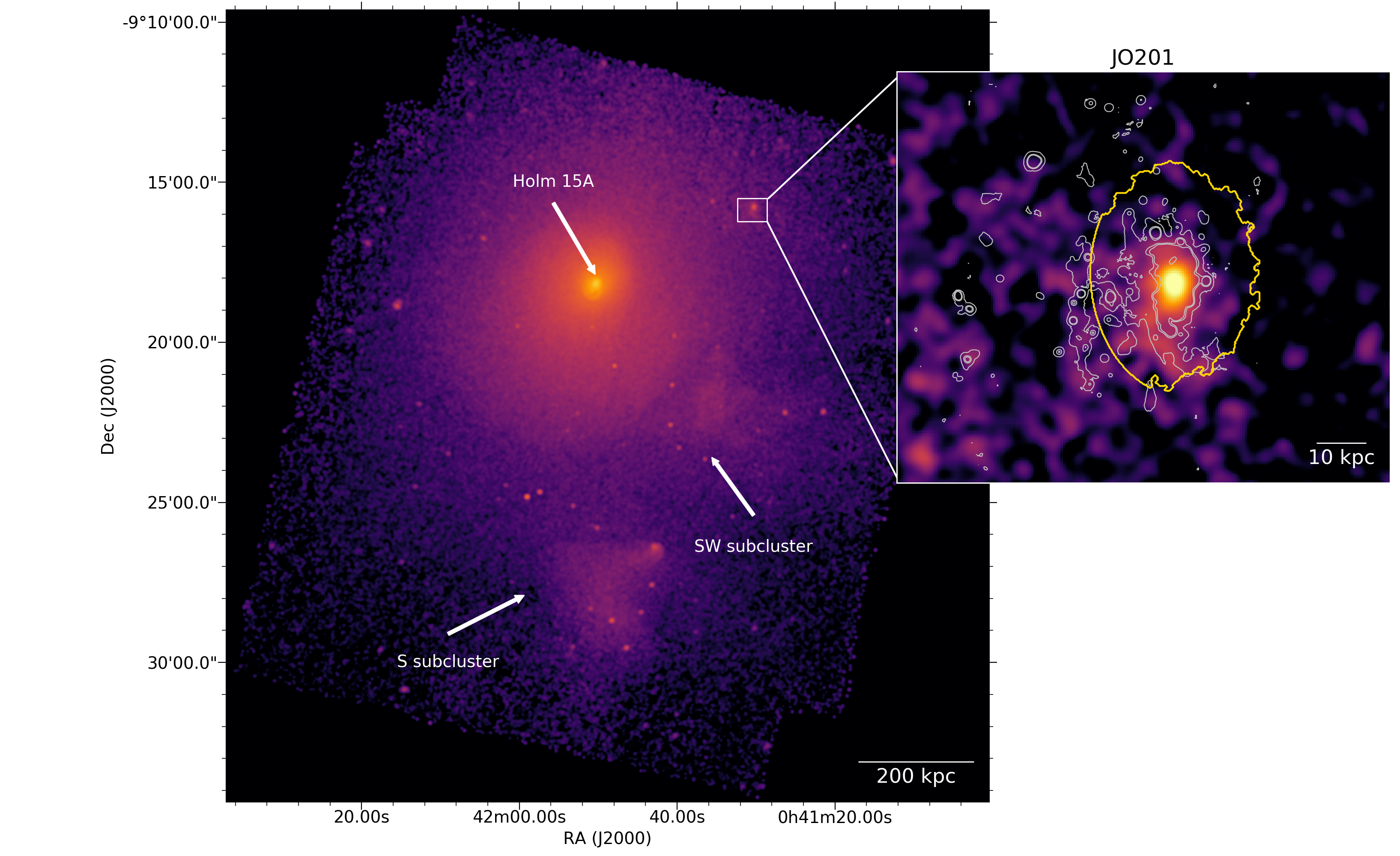}
      \caption{Exposure-corrected mosaic of the cluster Abell 85, Gaussian-smoothed with $\sigma$=2.48 arcsec, obtained by combining the five multi-chip Chandra observations and by filtering in the 0.5-2.0 keV band (see §\ref{chandarepro} for details). In the zoomed-in image is shown the galaxy JO201: white contours represent the H$\alpha$ emission with S/N$>$3 \citep{Bellhouse2019}, while the gold contour represents the stellar disk region \citep{Gulli2020}.}
         \label{mosaico}
   \end{figure*}

\subsection{JO201: the current state of the art}\label{JO201}
JO201 is a Seyfert 2 galaxy \citep[RA 00:41:30.325, Dec -09:15:45.96,][]{Bellhouse2017} located at a distance of 360 kpc north west (NW) from the BCG \citep[RA 00:41:50.54, Dec -09:18:13.07,][]{Edwards2016}; it is characterised by a stellar mass of $M_{s}= 3.55^{+1.24}_{-0.23} \cdot 10^{10}$ M$_{\odot}$ and a line of sight velocity of v$=3363.7$ km $\text{s}^{-1}$ with respect to the mean velocity of the cluster \citep{Bellhouse2017,Vulcani2018}. Its proximity to the centre of A85,  along with its very high velocity, make the stripping process extremely efficient. Referring to the main questions that the GASP project aims to answer \citep{Poggianti2017}, in this subsection we present both the results obtained from previous optical and radio studies of JO201 and the open questions. 
\begin{enumerate}
    \item For how long, where and why does gas removal occur? The motion of JO201 develops almost exclusively along the line of sight in the direction of the observer, providing a unique perspective on the stripping process. The orientation of the galaxy is indeed face-on and the interaction with the ICM occurs frontally. This allows to recognise a particular pattern in the distribution of the H$\alpha$ emission attributable to the unwinding of the spiral arms \citep{Bellhouse2020}. Observations revealed that the efficiency of the stripping is such that gas was removed even from the innermost part of the stellar disk. From N-body simulations it was estimated that the process has been ongoing for $\sim$ 0.6-1.2 Gyr and thus a long extension of the gas tail is expected \citep{Bellhouse2019}. H$\alpha$ observations highlight the presence of gas up to a distance of 50 kpc from the disk, but this value could be underestimated due to the particular orientation of JO201. For this reason, \cite{Bellhouse2019} resorted to simulations to investigate the motion and dynamics of a cloud of gas that is accelerated away from the galaxy, after being stripped by ram pressure, finding that the real length of the tail is probably $\sim$ 94 kpc;
    
    \item What is the impact of the RPS process on star formation? Previous studies found that JO201 has a total integrated SFR of $6 \pm 1$ M$_{\odot}$yr$^{-1}$, which is 0.4–0.5 dex above the main sequence of non-stripped disk galaxies \citep{Vulcani2018}, and shows a molecular gas content 4-5 times higher than normal galaxies of similar mass \citep{Moretti2020}. In particular, the neutral gas mass (HI) is 14 times lower than the molecular gas mass ($H_{\text{2}}$), which is in contrast with the behaviour of normal spiral galaxies. The hypothesis  suggested until now to interpret these features, is that ram pressure could be able to compress the gas, increasing the SF and converting HI into H$_{\text{2}}$ effectively \citep{Ramatsoku2020};
    
    \item What is the impact of the AGN activity on the RPS process? JO201 hosts an AGN whose activity influenced the central region, both ionising the surrounding gas \citep{Poggianti2017,Bellhouse2019, Radovich2019} and quenching the star formation \citep{George2019}. However, a detailed analysis of its X-ray emission has not been carried out yet.

\end{enumerate}
Therefore, it clearly appears that there are still several shortcomings in the understanding of the ongoing stripping process in JO201, regarding the nature of the interaction between the ISM and the ICM, and the properties of the AGN. In this paper, we aim at investigating the origin of the extended X-ray emission associated to JO201, by means of a dedicated re-analysis of the archival Chandra observations \citep[first presented by][]{Ichinohe2015}. In particular, we aim at discerning if the X-ray emission is linked to the intense ongoing star formation or to a further contribution caused by the RPS process and also at analysing the AGN, in order to understand the origin of its high-energy emission.

\section{Data analysis}

\subsection{Chandra data processing \label{chandarepro}}
We analysed archival Chandra observations of the cluster A85 consisting of five different exposures (ObsID 15173, 15174,16263, 16264, 904, see Table \ref{obs}) hosting the galaxy JO201. Each observation was obtained with the ACIS-I instrument in the VFAINT mode and the cluster emission extends over all the four chips. Data were downloaded from the \textit{Chandra Data Archive}\footnote{\url{https://cda.harvard.edu/chaser/}} and were reprocessed with the software package CIAO (version 4.11) and CALDB (version 4.8.3) to apply the latest calibration and to remove bad pixels and flares. The net exposure times obtained after this phase of data cleaning are summarised in Table \ref{oss}. We also improved the absolute astrometry identifying the point sources with the task \texttt{WAVDETECT} and cross-matching them with the optical catalogue USNO-A2.0\footnote{\url{http://tdc-www.harvard.edu/catalogs/ua2.html}}. 
For background subtraction we first identified and re-project the ACIS background files ({\ttfamily blanksky}) that matches our data; then we treated each chip of each {\ttfamily blanksky} separately, normalising them to the count rate of the corresponding chip of the source image in the 9-12 keV band.\\

For each observation we produced an image in units of counts (\texttt{counts/pixel}), an exposure map (\texttt{cm}$^2\cdot$\texttt{s}$\cdot$\texttt{counts/ photons}) and an image in units of flux (\texttt{photons/(cm}$^2\cdot$\texttt{s}$\cdot$\texttt{pixel})) resulting from the division between the image in units of counts and the exposure map. Combining the five images in units of flux and filtering data in the 0.5-2.0 keV band, we produced the exposure-corrected mosaic shown in Figure \ref{mosaico}. 

\begin{table}
\caption{Summary of the observations; the net exposure times are the values after cleaning.\label{obs}}             
\label{oss}      
\centering                          
\begin{tabular}{c c c c}        
\toprule                 
ObsID & Date & \begin{tabular}[c]{@{}c@{}}Net exposure \\ time (ks)\end{tabular} & PI \\    
\toprule                       \smallskip
   15173 & 2013-08-14 & 40.2  & S. Allen \\  \smallskip    
   15174 & 2013-08-09 & 38.2    & S. Allen \\ \smallskip  
   16263 & 2013-08-10 & 36.3     & S. Allen \\ \smallskip  
   16264 & 2013-08-17 & 34.3    & S. Allen \\ \smallskip
   904   & 2000-08-19 & 37.9   & C.L. Sarazin\\    
\bottomrule                                   
\end{tabular}
\end{table}  
   
\subsection{X-ray morphology}  
From the exposure-corrected mosaic (Figure \ref{mosaico}), it is possible to recognise the main features of the cluster A85 presented in §\ref{Abell85}. In particular, in the S and SW regions we identify the two sub-clusters that are falling towards the centre of A85. Despite the presence of these two mergers, the gas distribution follows a spherical symmetry elsewhere in the cluster. \\ 

To the NW of the BCG, in correspondence to the position of JO201, it is possible to observe an extended X-ray emission. Specifically, we distinguish a central region related to the centre of the galaxy disk including the AGN and a southern, arcuate region resembling the unwinding of the spiral arm \citep{Bellhouse2020}. The X-ray emission is less extended than the optical disk, but, as in the case of JW100 (see §\ref{introduzione}), it coincides spatially with the H$\alpha$ emission (see also Figure \ref{mascherinaH}).

\subsection{Point-to-point analysis}
In order to unfold the connection between the X-ray emitting plasma and the galactic warm gas, we carried out a point-to-point study of the spatial correlation between the two respective emissions, i.e. X-rays and H$\alpha$.\\

We combined the Chandra image with MUSE H$\alpha$ cubes and emission fraction of the diffuse ionised gas (DIG) maps \citep[][]{Reynolds1985, Reynolds1999, Reynolds1999b, Tomicic2021}. The H$\alpha$ observations of JO201 were taken as part of the GASP program with the MUSE IFU instrument at the Very Large Telescope and processed following the standard procedure for GASP galaxies \citep[][]{Poggianti2017}. The observations was further corrected for extinction by dust within the Milky Way by estimating the contribution using the NASA/IRSA infrared science archive. In order to extract emission line fluxes, the cube was analysed using the IDL code \textsc{kubeviz} \citep{Fossati2016} after smoothing spatially with a 5x5 kernel. \textsc{kubeviz} fits a series of gaussian components to selected emission lines in each spaxel of the IFU data. The resulting map of H$\alpha$ emission line flux was used in the following analysis. 

The maps of emission lines corrected for attenuation, were calculated assuming the intrinsic Balmer line ratio of $\rm H\alpha/H\beta=2.86$, a ionised gas temperature of $\rm T\approx10^4$ K  and case B recombination \citep[][]{Osterbrock1992}. Finally, we discriminated the ionisation source of the H$\alpha$ emission (SF, LINER or AGN) according to the [OI] diagnostic BPT diagram \citep{Baldwin1981, Veilleux1987}. As reported in \citet[][]{Poggianti2019b}, at distances of 10-50 kpc from the galaxy centre, JO201 shows [OI]/H$\alpha$ ratios much higher than what observed in SF regions. The other BPT line ratios ([NII]/H$\alpha$ and [SII]/H$\alpha$) are instead closer to values measured in SF regions. Since the LINER-like emission is extended and not concentrated in the nucleus, in this work we refer to these regions as "LIERs" rather than LINERs, following \cite{Belfiore2016}.\\

Unlike the dense ionised gas in HII regions, the DIG component of the ISM  has low densities  (n $ \sim 10^{-1}$ cm$^{-3}$),  high temperatures (higher than $\sim10^4$ K), and can  reach large spatial distances from the HII regions \citep[up to 1-2 kpc,][]{Haffner2009}. The DIG exhibits lower H$\alpha$ surface density and higher $\rm [S\textsc{ii}]/H\alpha$ line ratio compared to the dense gas emission. Fitting the anti-correlation between $\rm [S\textsc{ii}]/H\alpha$ ratio (accounted for radial gas-phase metallicity gradient) and the extinction corrected H$\alpha$ surface brightness, $\rm \Sigma H\alpha_{corr}$, we derived the maps of fraction of H$\alpha$ emission due to DIG (labelled hereafter as $\rm C_{DIG}$) through JO201. The details of  the DIG fraction estimation are described in \citet{Tomicic2021}.  \\

\begin{figure}
    \centering
    \includegraphics[width=\linewidth]{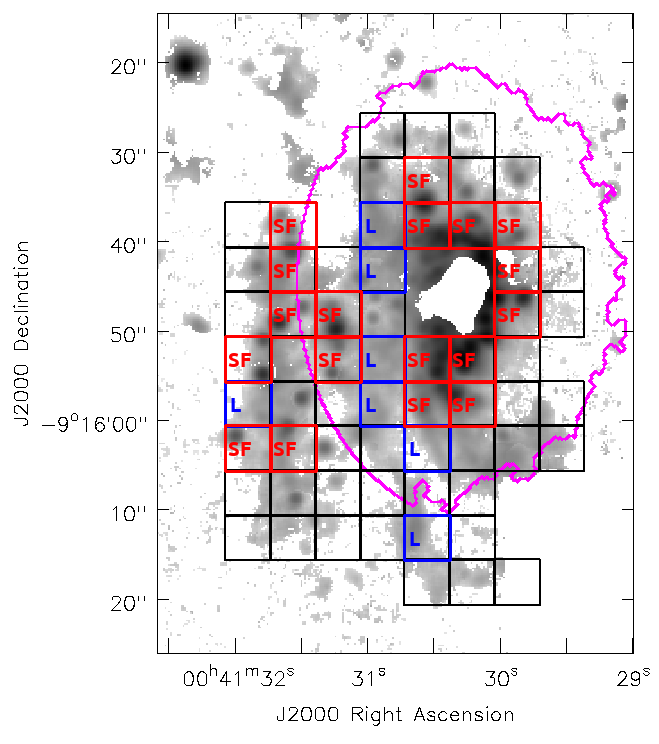}
    \caption{MUSE H$\alpha$ image with the stellar disk (magenta) and the sampling grid (black) on top, including the SF- and LIER-dominated regions in, respectively, red and blue. \label{griglia}}
\end{figure}

The following point-to-point analysis was carried out with the PT-REX code \citep[][Ignesti in preparation,]{Ignesti2020} which is based on the Common Astronomy Software Application package\footnote{\url{https://casa.nrao.edu/}} (CASA v. 6.0). For sampling the H$\alpha$ emission, we resort to a grid composed of 5 arcsecond-large cells, corresponding to $\sim$5 kpc. This sampling-scale allows us to reconcile the X-ray S/N of each cell (which increases with the size of the cell) with the number of sampled points (which decreases with the size of the cell). The resulting sampling scale is larger than the typical size of SF-blobs ($\sim$1-2 kpc) evidenced by the H$\alpha$ emission in MUSE, and therefore these substructures are sub-sampled in our analysis. In order to optimise the signal of the galactic emission, we considered only the regions of the H$\alpha$ image with S/N$>$3 and we excluded all those pixels that, according to the BPT diagram, are characterised by the AGN-related emission. Finally, we used the stellar disk mask (Figure \ref{griglia}) to discriminate the cells within the stellar disk (i.e. those that overlap with the stellar disk mask for $>50\%$ of the cell area) from those located over the tail. \\

We further refined our sampling by identifying those cells with $>50\%$ SF-related emission (red) and $>50\,\%$ LIER-related emission (blue). However, since it was not possible to determine a BPT classification for every spaxel, which is only accepted if all the 4 lines invoked have S/N$>$3, we discern the dominant process only in the brightest regions of the galaxy. For this reason, the BPT-spatial correlation analysis does not cover the whole galaxy, but is still able to provide insights of the general trends.

The sampling grid obtained is reported in Figure \ref{griglia} and is composed of 63 cells, of which 18 were classified as SF-dominated and 7 as LIER-dominated. We note that, considering the resolution of the image, the large number of sampling cells assure us that our results are not biased by the choice of the geometry of the sampling grid \citep[we refer to][for a more detailed discussion of this issue]{Ignesti2020}.\\

For each cell, we measured both the total H$\alpha$ surface brightness and the DIG fraction, which we estimated as the total DIG flux density (obtained by multiplying $\rm C_{DIG}$ and  H$\alpha$ in each spaxel) of each cell divided by the H$\alpha$ flux density, and compared them to the X-ray surface brightness. The correlations were evaluated not only for the whole population, but also for the SF and LIERs regions separately (Figure \ref{ptp}). In particular, using the orthogonal BCES algorithm, which is able to account for the internal scatter of the data, \citep[][]{akritas1996}, we fitted the data with a power-law relation $I_{H\alpha}\text{, DIG fraction}\propto I_\text{X}^k$. Finally, in order to estimate the strength of the linear correlations, we computed the Pearson and Spearman correlation ranks ($\rho_\text{P}$ and $\rho_\text{S}$), which are reported in the legend of Figure \ref{ptp}, the p-values ($p_\text{P}$ and $p_\text{S}$), and the correspondent correlation probabilities ($P_\text{c,P}$ and $P_\text{c,S}$). The best-fitting indexes, correlation ranks, p-values and probabilities are reported in Table \ref{ptp-tab}.

\subsection{Spectral Analysis}
\subsubsection{Spectra extraction procedure}\label{spectraext}

\begin{figure*}
        
        \begin{subfloat}{}
           \centering
            \includegraphics[width=3.in]{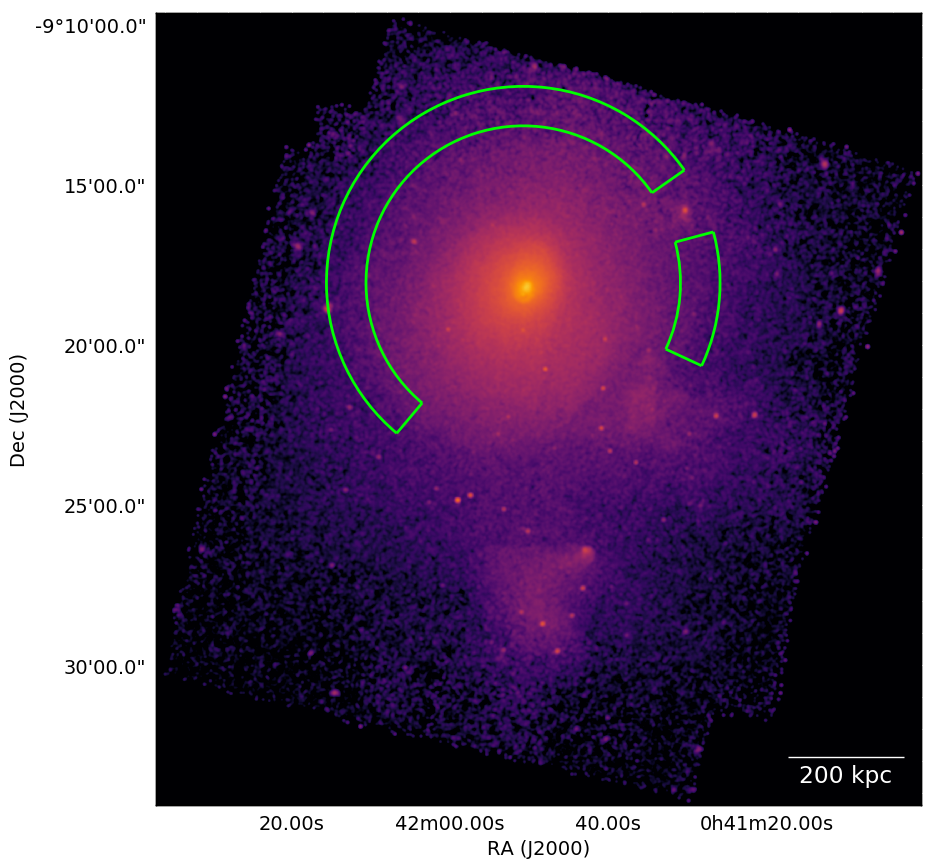}
         \end{subfloat}
        \hfill
        \begin{subfloat}{}
           \centering
             \raisebox{0.1\height}{\includegraphics[width=3.in]{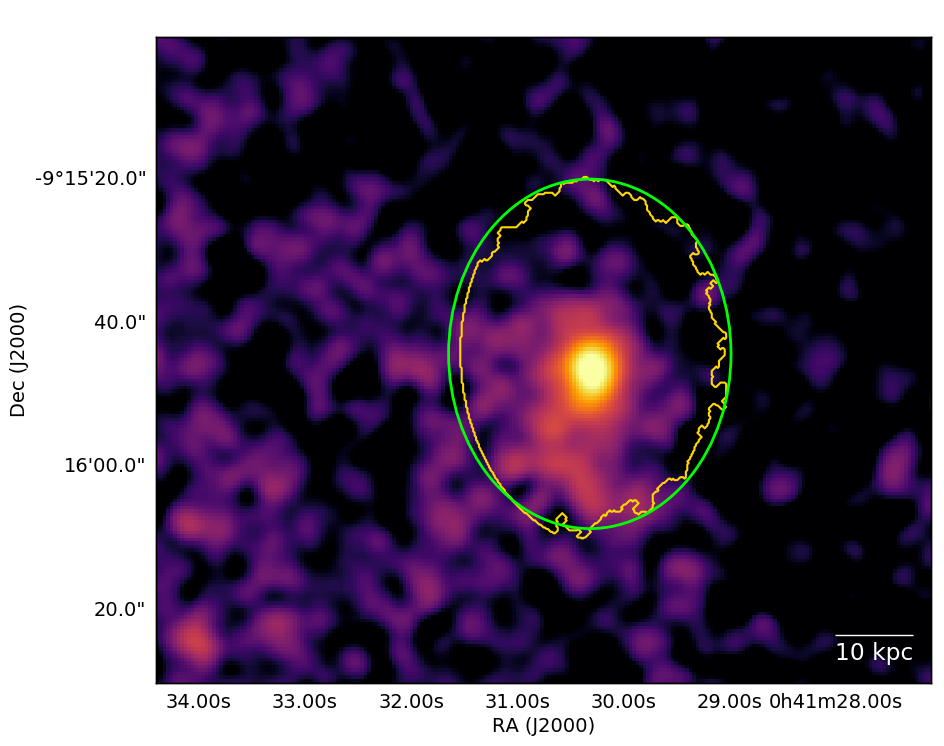}}
      \end{subfloat} 
  \caption{Left panel: exposure corrected mosaic of the cluster A85 in the 0.5-2.0 keV, in green is reported the region used for the ICM spectra extraction. Right panel: zoomed-in image of JO201, in green is represented the region used to extract spectra from the stellar disk mask, while in gold is reported the contour of the stellar disk \citep{Gulli2020} used as reference. }\label{maschere}
 \end{figure*}

The spectral analysis of JO201 was carried out with the software package XSPEC \citep[version 12.10.1f,][]{Arnaud1996}. For each observation, we masked all the point sources identified by the \texttt{WAVDETECT} in §\ref{chandarepro} except the AGN of JO201. In order to study the nature of the interaction between the ICM and the galaxy, we followed the approach proposed in \cite{Poggianti2019} by considering three regions. The first two relate to the galaxy: one (Figure \ref{mascherinaH}) follows the contours of the H$\alpha$ emission \citep{Bellhouse2019} and resembles the X-ray emission; the other one (Figure \ref{maschere}, right panel) traces the contours of the stellar disk and was chosen to compare the SFR estimated from the X-ray luminosity to the SFR obtained in a previous study on the H$\alpha$ emission in the disk \citep{Vulcani2018,Gulli2020}. The third region (Figure \ref{maschere}, left panel) was used to determine the properties of the ICM at the same clustercentric distance of JO201. Under the assumption of the spherical symmetry of the plasma distribution in the northern region of the cluster in which JO201 is moving, we selected an annular region excluding the southern-west part of Abell 85, where the merger with the sub-clusters S and SW is ongoing (see §\ref{Abell85}).\\

In order to generate the appropriate response matrices, spectra were extracted separately from each observation (or from each chip of each observation in the case of the annular region) and were fitted jointly in the 0.5-7.0 keV band after background subtraction. In the case of the ICM analysis, background spectra were extracted from \texttt{blanksky} files in the same annular region described above. For the galaxy analysis, instead, we tested different extraction procedures which we summarise in Table \ref{casistica}. On the one hand, we considered two different backgrounds, i.e.: case 1) normalised CALDB {\ttfamily blanksky} files (see §\ref{chandarepro}), or case 2) local background extracted in an annular region at the same distance and on the same chip of the galaxy. The procedure followed in this latter case allows to subtract the contribution of the ICM located along the line of sight, that has not been modelled in the fit. On the other hand, we treated the AGN with two different approaches: case A) including it in the spectrum and modelling in the fit, or case B) excluding it during the spectrum extraction, by masking an elliptical region of the same size of the Chandra PSF at that distance from the pointing centre.

\subsubsection{Spectral models \label{models}}
\begin{table*}
\caption{Configuration used for the spectra extraction and background/AGN components added to the component that models the galactic emission (i.e. \texttt{apec}, \texttt{cemekl} or \texttt{mkcflow}) }\label{casistica}
\centering
\hspace{-1.2cm}
\begin{tabular}{@{}lcc@{}}
\toprule
              & \begin{tabular}[c]{@{}c@{}}AGN included \\ (case A) \end{tabular}                                                & \begin{tabular}[c]{@{}c@{}}AGN excluded \\ (case B) \end{tabular}                                            \\ \midrule
\begin{tabular}[c]{@{}l@{}}{\ttfamily blanksky}\\ (case 1) \end{tabular}     & \begin{tabular}[c]{@{}c@{}}\texttt{apec+pow}\end{tabular} & \begin{tabular}[c]{@{}c@{}}\texttt{apec}\end{tabular} \\
\begin{tabular}[c]{@{}l@{}}Local background \\ (case 2) \end{tabular}    & \begin{tabular}[c]{@{}c@{}}  \texttt{pow}\end{tabular} &\begin{tabular}[c]{@{}c@{}}   - \end{tabular}                                           \\  \bottomrule
\end{tabular}

\end{table*}

Following the approach presented in \cite{Poggianti2019}, we first investigated the properties of the environment surrounding JO201 to model the contribution of the ICM to the galactic X-ray emission. We fitted the ICM spectra with an absorbed thermal model, \texttt{tbabs$\cdot$apec}. The \texttt{tbabs} component represents the galactic absorption, and its column density parameter, n$_H$=2.7$\cdot$10$^{20}$cm$^{-2}$ was fixed; the \texttt{apec} component instead, describes the emission of a single temperature plasma and its temperature, metallicity and normalisation parameter were let free to vary. The abundance tables used for the entire spectral analysis refer to \cite{Asplund2009}.\\

We then focused on the galaxy to investigate the origin of its extended high-energy emission. In particular, we explore four possible scenarios, which attribute the emission to: i) the ongoing  SF, ii) the stripping of a galactic hot halo that the galaxy may have possessed, iii) either the cooling of the ICM onto the galaxy, or the heating of the ISM through shocks and conduction, iv) the mixing of the ICM cooling and ISM heating.\\

To discern which of these hypothesis is correct, we carried out the spectral analysis of JO201. In particular, we tested complex models made of a galactic component chosen among the following one:
\begin{itemize}
    \item an \texttt{apec} model, which describes the emission of a single temperature plasma and can represent the SF, the hot halo scenarios, and, in a first approximation,  the ICM cooling- or the ISM heating-dominated scenarios; 
    \item a \texttt{cemekl} model, describing the emission of a multi-phase and multi-temperature plasma where the emission measure, EM= $\int n_{\text{e}} n_{\text{H}} dV$, scales with temperature as $EM \propto T^{\alpha}$ and T has an upper limit, $T_{\text{MAX}}$. This model can represents both the hot halo and the ICM-ISM mixing and from the index $\alpha$ it is possible to derive the mass-weighted temperature of the plasma as $T_{mw}=T_\text{MAX}(1+\alpha)/(2+\alpha)$ \citep[][]{Sun2010};
    \item  a \texttt{mkcflow} model, which is usually used to describe the thermal cooling of a gas from a maximal, T$_{\text{MAX}}$, to a minimal temperature, T$_{\text{min}}$. This model returns the mass accretion rate parameter ($\dot{\text{M}}$) and, in our picture, can represent the ICM cooling scenario \citep{Mushu1988}. 
\end{itemize}
We include also an additional component that, depending on the extraction procedure used (see §\ref{spectraext}), models either the ICM along the line of sight or the AGN emission. In particular, we add to the galactic component: 
 \begin{itemize}
 \item an \texttt{apec} component when the {\ttfamily blanksky} was used and the AGN was excluded (case 1B). This component takes into account the ICM emission along the line of sight and its parameter were set to the value determined from the ICM analysis;
 \item both an \texttt{apec} and a \texttt{powerlaw} component when the {\ttfamily blanksky} was used and the AGN was included (case 1A). We assume in fact, that the emission of this central source can be modelled with a power law;
 \item a \texttt{powerlaw} component to model the emission of the AGN when it was included and the local background was used (case 2A);
\item no component when the AGN was excluded and the local background was used (case 2B).
 \end{itemize}
 In each test we account for the galactic absorption by including a {\ttfamily tbabs} component with a column density of n$_{\text{H}}$=2.7$\cdot$10$^{20}$cm$^{-2}$ \citep{Kaberla2005}. In Table \ref{casistica} we summarise the four configurations used to extract spectra and the relative components added to the galaxy emission model.

\section{Results} \label{risultati}

\subsection{Spatial correlation}

\begin{figure*}
\includegraphics[width=\linewidth]{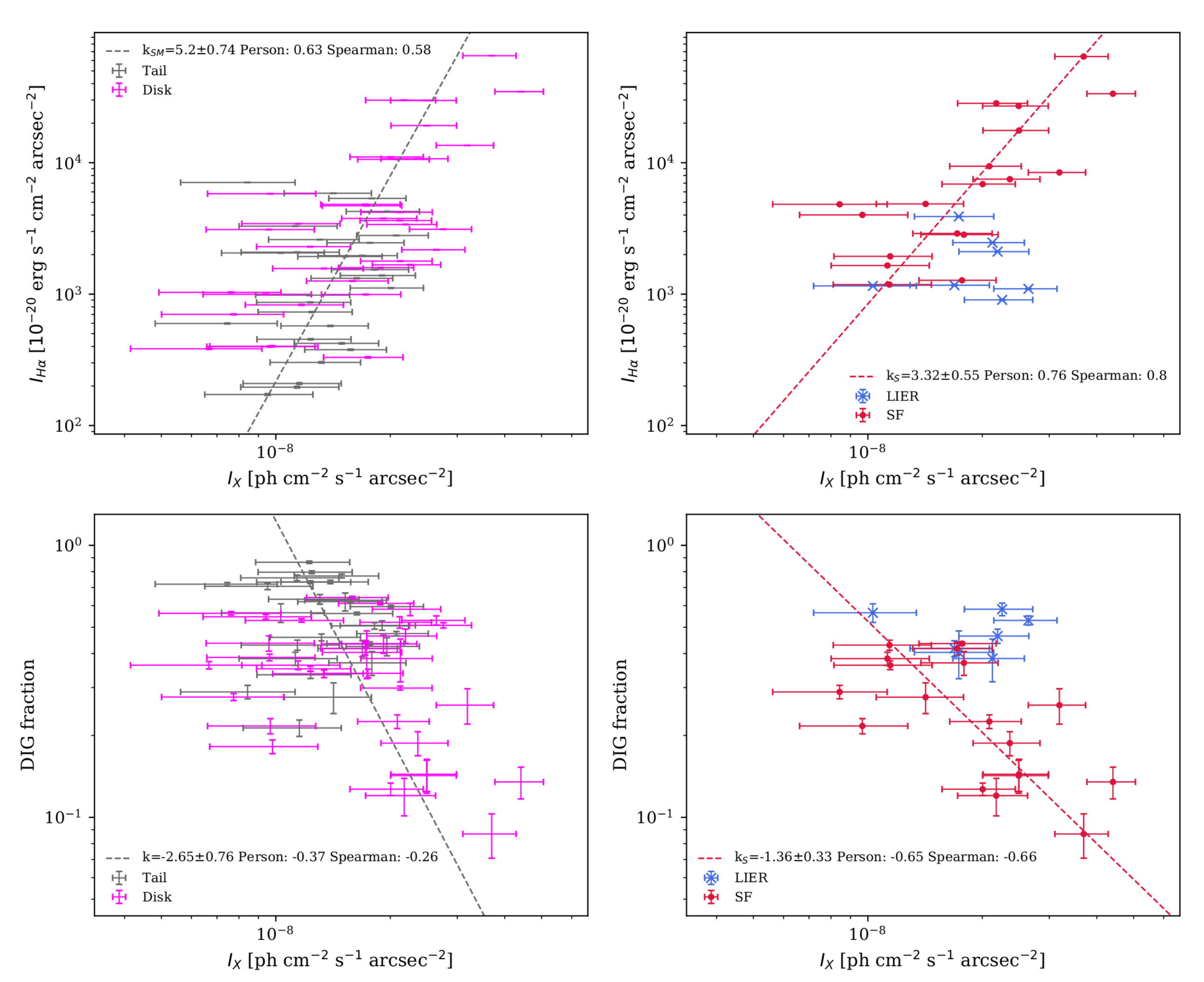}
\caption{\label{ptp} Resulting $I_{H\alpha}$ (top) and DIG fraction (bottom) correlations obtained with the grid presented in Figure \ref{griglia}. In the left panels we report the distributions obtained by using every cells, where we report in magenta the cells within the disk and in black those in the tail. In the right panels are shown the spatial correlations of the SF (red) and LIER (blue) regions. We best fit indexes are reported in the legends and summarised in Table \ref{ptp-tab}.}
\end{figure*}

In Figure \ref{ptp} and in Table \ref{ptp-tab} we report the results of the point-to-point analysis carried out with the grid presented in Figure \ref{griglia}. The majority of regions are characterised by lower H$\alpha$ and X-ray surface brightness and higher DIG fraction, whereas the brightest spots, which are located in the innermost part of the disk, show a lower DIG fraction. The BPT-based sampling highlights that the SF-dominated regions have, generally, an higher $I_\text{H$\alpha$}$ and lower DIG fraction with respect to the LIER-dominated regions. \\

We observe a positive correlation between H$\alpha$ and X-ray surface brightness (Figure \ref{ptp}, top-left panel), and a putative anti-correlation between DIG fraction and X-ray surface brightness (Figure \ref{ptp}, bottom-left panel). The two correlations become more evident when we limited our analysis to the SF-dominated regions. On the contrary, LIER-dominated regions do not show any evident correlation (Figure \ref{ptp}, right panels). \\

\begin{table*}
\caption{ Results of the point-to-point analysis presented in Figure \ref{ptp}. From left to right: sampled points; Index of the best-fit power-law; Pearson rank; Pearson p-value; Pearson correlation probability; Spearman rank; Spearman p-value; Spearman correlation probability.}
    \label{ptp-tab}
    \centering
    \begin{tabular}{lccccccc}
    \toprule
    &$k$&$\rho_\text{P}$&$p_\text{P}$&$P_\text{c,P}$&$\rho_\text{S}$&$p_\text{S}$&$P_\text{c,S}$\\
    \midrule
     \multicolumn{8}{c}{$I_{H\alpha}$ vs $I_\text{X}$}\\
     \midrule
    Total&5.17$\pm$0.73&0.63&3.32$\times10^{-8}$&0.99&0.58&7.99$\times10^{-7}$&0.99\\
    SF&3.32$\pm$0.55&0.76&2.86$\times10^{-4}$&0.99&0.80&5.82$\times10^{-5}$&0.99\\
    LIER&none&0.02&0.96&0.04&-0.39&0.38&0.62\\
    \midrule
     \multicolumn{8}{c}{DIG fraction vs $I_\text{X}$}\\
     \midrule
    Total&-2.58$\pm$0.74&-0.37&2.86$\times10^{-3}$&0.99&-0.26&3.95$\times10^{-2}$&0.96\\
    SF&-1.29$\pm$0.30&-0.65&3.25$\times10^{-3}$&0.99&-0.66&3.16$\times10^{-3}$&0.99\\
    LIER&none&-0.14&0.82&0.18&0.25&0.59&0.41\\
    \bottomrule
    \end{tabular}

\end{table*}


\subsection{Results of the spectral analysis}
In this subsection we report the results obtained by the spectral analysis of both the AGN and the galactic emission.
\subsubsection{AGN}
The central source in JO201 has a shape consistent with a point source at a distance of $\sim$5 arcmin from the aimpoint. We extract the spectrum from a circular region of 4 
arcsec enclosing the central source, and we use the surrounding annulus as a background that includes the underlying cluster (A85) background, the instrumental background and the galaxy diffuse background. The resulting number of net counts is not enough to allow a thorough spectral analysis: with around 100 net counts, the extracted spectrum is well fitted by a power law, absorbed by the Galactic n$_{\text{H}}$. However, the resulting slope, $\Gamma$ = 2.7$\pm$ 0.2 is steeper than usual in AGNs. A more complex model is therefore adjusted to the data: a \texttt{pexrav} with the addition of a Gaussian line. The model describes the reflection spectrum off the walls of the torus surrounding the AGN. The match is good with an intrinsic "canonical" $\Gamma$ = 1.7 and a high degree of reflection, as in a Compton Thick AGN. The energy of the Gaussian line is consistent with a 6.4 keV K$\alpha$Fe emission redshifted at the z=0.055 of the galaxy. The resulting luminosity is L$_X^{0.5-10 keV}$=2.7 $\times 10^{41}$ erg s$^{-1}$ \footnote{We note that this is lower than reported in \citet[][]{Poggianti2017b} (L$_X^{0.3-8 keV}$=7.3 $\times 10^{41}$ erg s$^{-1}$) due to the different spectral model we adopted.}. However, the intrinsic luminosity could be at least a factor of a hundred higher, given the torus absorption. This is also in accordance with the optical classification of the source as a Type 2 \citep{Radovich2019}. A higher statistics, or observations at higher energies would be needed to derive precise parameters for the source. 

\subsubsection{Galaxy}\label{discos}
Our spectral analysis of the X-ray emission associated to JO201 did not reveal significant discrepancies between the four procedures tested for the spectra extraction (see §\ref{spectraext}).
In this section we report the tests performed with the configuration that requires the simplest model and ensures the higher number of counts, i.e. 1B ({\ttfamily blanksky} and AGN subtraction). The spectral results obtained with the H$\alpha$ and the stellar disk mask are consistent in terms of both the best-fit parameters and the $\chi^2$ values. For this reason, aiming to realise a comparison between the SFR estimated from the X-ray luminosity and the SFR found in \citep[][]{Vulcani2018} from the H$\alpha$ emission in the disk, we choose to present the results obtained with the stellar disk mask, whereas those obtained with the H$\alpha$ mask are presented in the Appendix \ref{risultha}. \\

Following the approach described in Section \ref{modelli}, we first estimated the properties of the surrounding ICM. In agreement with previous results of \cite{Ichinohe2015} (see § \ref{Abell85}), we found a temperature of kT=$7.1 \pm 0.2$ keV and a metallicity of Z=$0.21 \pm 0.04$ Z$_{\odot}$. Based on the temperature, we estimated a local speed of sound $c_s\simeq520\sqrt{k\text{T}/\text{(1 keV)}}\simeq1376$ km s$^{-1}$, which entails a galactic Mach number $\mathcal{M}\simeq2.4$ .\\

We then carried out the modelling of the X-ray emission from the region of the stellar disk, following the procedure described in section \ref{models}. In particular, we built our model by adding to the galactic component (\texttt{apec}, \texttt{cemekl} or \texttt{mkcflow}) an \texttt{apec} component that takes into account the ICM emission along the line of sight (1B configuration in Table \ref{casistica}); we report the results of the fits in Table \ref{modelli}.\\

Our criterion to attest the validity of a model, was based not only on the values of the $\chi^2$ obtained, but also on the values of the temperature, T, and metallicity, Z revealed by the fits. Thus, we have found a temperature of kT=0.79$^{+0.08}_{-0.08}$ keV for the model 1) \texttt{apec+apec}, kT$_{\text{MAX}}$=1.2$^{+0.4}_{-0.2} $ keV for the model 3) \texttt{apec+mkcflow}, and kT$_{\text{MAX}}$=1.1$^{+0.5}_{-0.2} $ keV for the model 4) \texttt{apec+mkcflow}. In all these three cases, the values obtained are lower than the ICM temperature, sign that the emitting plasma is in an intermediate phase between the one of the ICM and of the ISM. For the model 2) \texttt{apec+cemekl}, instead, we noticed that by leaving the temperature parameter free to vary it was not possible to obtain a good fit of the data. Since the \texttt{cemekl} model describes the emission of a multi-phase and multi-temperature plasma, we supposed that the maximal value that the temperature can assume is the value of the ICM temperature, and for this reason we fixed $T_{\text{MAX}}=7.1$ keV. The best-fit $\alpha=0.01^{+0.10}_{-0.01}$ would entail a $T_{mw}\simeq3.5$ keV and an almost uniform EM of the plasma.\\

Regarding metallicity, we note that, when it was set free (models 1 and 4), the Z values obtained are lower than that of the ICM. We considered these values implausible, because the X-ray emitting plasma is either the ICM, or the ISM or a mixing of these two components and thus its metallicity is expected to vary between the values of the ICM metallicity (Z$_{\text{ICM}}$=0.21 Z$_{\odot}$, see § \ref{models}) and  of the ISM metallicity (typical values are Z$\sim$1 Z$_{\odot}$ ). In the other two cases (models 2 and 3) it was not possible to obtain a good fit of the data by letting this parameter free, for this reason we fixed it to the value of the ICM. \\

Under a statistical point of view, the four models are indistinguishable because they fit equally well the observations,  as it is possible to notice from the values of the $\chi^2_{\text{R}}$.

\begin{table*}
\caption{Fit results for spectra extracted from the ICM region ("Control region", Figure \ref{mosaico}) and from the stellar disk mask ("JO201", Figure \ref{maschere}, right panel). For this latter region, we resort to the 1B extraction configuration, which models the ICM emission with an additional \texttt{apec} component (see §\ref{models}). The best-fit parameters shown in the third column refer only to the galactic component indicated in the second column (namely 1) \texttt{apec}, 2)  \texttt{cemekl}, 3) and 4) \texttt{mkcflow}).} \label{modelli}  \
\hspace{-1.5 cm}
\footnotesize
\begin{tabular}{@{}llcc@{}}
\toprule
                                   & Model  & Parameters\tnote{a}  & \begin{tabular}[c]{@{}c@{}} $\chi^2$/dof \smallskip \\ $\chi^2_{\text{R}}$  \end{tabular}   \\ \toprule
Control region (ICM)                & \begin{tabular}[c]{@{}l@{}}0)\texttt{tbabs$\cdot$apec}\end{tabular} & \begin{tabular}[c]{@{}c@{}} kT=7.1$\pm 0.2$ , Z=0.21$\pm 0.04 $ \smallskip \\ F=(8.65$\pm$0.17)$\cdot$10$^{-13}$,   L=(2.73 $\pm$ 0.08)$\cdot$10$^{42}$ \end{tabular}  &\begin{tabular}[c]{@{}c@{}}1694.7/1617 \smallskip \\ 1.048\end{tabular}  \\ \midrule

\multirow{10}{*}{JO201}        & \begin{tabular}[c]{@{}l@{}}1) \texttt{tbabs$\cdot$(apec+apec)}\end{tabular} & \begin{tabular}[c]{@{}c@{}} kT=0.79$^{+0.08}_{-0.08}$ , Z=0.08$^{+0.08}_{-0.04} $  \smallskip \\ F=(4.1$\pm$0.5)$\cdot$10$^{-14}$, L=(1.9 $\pm$ 0.3)$\cdot$10$^{41}$ \end{tabular}  &  \begin{tabular}[c]{@{}c@{}}43.18/47 \smallskip \\ 0.9187\end{tabular} \\ \cmidrule{2-4} 

&\begin{tabular}[c]{@{}l@{}} 2) \texttt{tbabs$\cdot$(apec+cemekl)} \\ Z=0.21 and \\ kT$_{\text{MAX}}=7.1$ keV (fixed)\end{tabular} & \begin{tabular}[c]{@{}c@{}} $\alpha$=0.01$^{+0.10}_{-0.01}$ \\ \smallskip F=(9.4$\pm$0.8)$\cdot$10$^{-14}$, L=(4.5$\pm$ 0.4)$\cdot$10$^{41}$ \end{tabular} & \begin{tabular}[c]{@{}c@{}}40.76/48 \smallskip \\ 0.8492\end{tabular} \\ \cmidrule{2-4} 

& \begin{tabular}[c]{@{}l@{}}3) \texttt{tbabs$\cdot$(apec+mkcflow)}\\ Z=0.21 (fixed)\end{tabular} & \begin{tabular}[c]{@{}c@{}} kT$_{\text{min}}$=0.08$^{+0.27}_{-0.08}$ , kT$_{\text{MAX}}$=1.2$^{+0.4}_{-0.2} $  \smallskip \\ $\dot{M}=1.7_{-0.6}^{+0.4}$  \smallskip \\ F=(4.4$\pm$0.6)$\cdot$10$^{-14}$, L=(1.9 $\pm$ 0.3)$\cdot$10$^{41}$ \end{tabular}  & \begin{tabular}[c]{@{}c@{}}40.07/47 \smallskip \\  0.8526\end{tabular} \\ \cmidrule{2-4}
                                   & \begin{tabular}[c]{@{}l@{}}4) \texttt{tbabs$\cdot$(apec+mkcflow)}\\ \end{tabular} & \begin{tabular}[c]{@{}c@{}} kT$_{\text{min}}$=0.08$^{+0.35}_{-0.08}$ , kT$_{\text{MAX}}$=1.1$^{+0.5}_{-0.2} $ \smallskip \\ $\dot{M}=2.1_{-1.1}^{+3.5}$. Z=0.13$^{+0.27}_{-0.09}$ \smallskip \\ F=(4.3$\pm$0.5)$\cdot$10$^{-14}$, L=(3.3 $\pm$ 0.4)$\cdot$10$^{41}$ \end{tabular}  & \begin{tabular}[c]{@{}c@{}}39.78/46 \smallskip \\ 0.8609\end{tabular} \\ \bottomrule 
\multicolumn{4}{l}{%
  \begin{minipage}{18cm}%
  \medskip
     \textbf{Notes}: The unit measure used are: keV for temperature (kT), Z$_{\odot}$ for metallicity (Z), M$_{\odot}$ yr$^{-1}$ for the mass accretion rate ($\dot{\text{M}}$), erg s$^{\text{-1}}$ cm $^{\text{-2}}$ for flux (F) and erg s$^{\text{-1}}$ for luminosity (L). The unabsorbed flux and luminosity refer only to the galactic component and were measured in the 0.5-10.0 keV energy band.%
  \end{minipage}%
}
           
\end{tabular}

\end{table*}

\section{Discussion}

\subsection{Origin of the X-ray emission}

The correlation between X-ray and H$\alpha$ surface brightness (Figure \ref{ptp}, top panels) indicates that the galactic X-ray emitting medium closely follows the spatial distribution of the ISM. This result suggests that the X-ray emission arises from local, small scale  processes taking place in the close proximity of the ISM. A similar behaviour was also observed both in the disk of isolated galaxies \citep{Ranalli2003, Symeonidis2011, Mineo2014} and in previous studies of \citet[][]{Sun2010} and \citet[][]{Poggianti2019} on jellyfish galaxies. In the first case, given the lack of a hot surrounding ICM, the correlation was interpreted as a consequence of recent star forming processes. In the second case, since JW100 and ESO-137-001 are located in a cluster environment and show an X-ray luminosity higher than what expected from star formation, it was suggested that the high-energy emission arises from the local interplay between the ICM and the ISM and that it could be produced by the interface between the ICM and the ISM. \\

Starting from these two different scenarios we investigated the contribution of the SF to the X-ray luminosity of JO201, following the approach presented in \cite{Poggianti2019}. In presence of SF indeed, the X-ray luminosity, of a galaxy is usually dominated by the contribution of both high mass X-ray binaries, whose characteristic lifetime is $10^7$ yr, and hot ISM, which is ionised by supernovae and massive stars \citep[e.g.,][]{Mineo2014}.Since the sum of these contributions correlates with other SFR indicators, many relations have been developed to link the $L_{\text{X}}$ of a galaxy to its SFR (and vice versa).\\
Using the $L_{\text{X}}$-SFR calibration of \cite{Mineo2014} converted from a Salpeter to a Chabrier IMF and from 0.5-8 keV to 0.5-10 keV assuming a factor 1.11:

\begin{equation}
     L_{X, 0.5-10.0 keV}=7.6 \cdot 10^{39} \, \text{SFR} \, \, \, \text{erg s}^{-1}
     \end{equation}
     
where the SFR is in unit of M$_{\odot}$yr$^{-1}$, we estimated a $L_{\text{X}}\sim 3.8 \cdot 10^{40}$ erg s$^{-1}$, which is lower by one order of magnitude than the values found by our models (see Table \ref{modelli}).\\

From this result, we conclude that the SF alone is not able to produce the observed X-ray luminosity and that an additional contribution is necessary, likely coming from the ongoing stripping process. 
This discrepancy between predicted and observed X-ray luminosity has been observed also for JW100 \citep[by a factor of $4-10$][]{Poggianti2019}, thus we suggest that it could be a common feature among jellyfish galaxies with extended X-ray emission. Further studies in this direction are now necessary to test our speculations.\\

The point-to-point analysis also revealed that the SF regions are generally above the LIERs ones in the H$\alpha$-X plot (Figure \ref{ptp}). An alternative view on these results is that, for similar values of $I_{H\alpha}$ or DIG fraction, the LIER regions show higher values of $I_\text{X}$ than SF regions. This latter feature is similar to what was previously observed in JW100 \citep[][]{Poggianti2019}. We also note that the main correlations are driven by the SF-dominated regions, whereas the LIER regions, which are present only in the low H$\alpha$ brightness or high-DIG fraction degree part of the plots, are less correlated. \\

Concerning the connection between DIG fraction and X-ray emission, we observe that LIER regions, which are located mostly along the stripped arm (Figure \ref{griglia}), show a generally higher DIG fraction. This behaviour suggests that the DIG regions in stripped tail may present more LIER features compared to the dense gas in the stellar disk, which is SF-dominated. Finally, the putative anti-correlation between the DIG fraction and $I_\text{X}$ could be explained by the fact that the DIG dominated regions typically have lower H$\alpha$ flux compared to the dense gas dominated regions \citep[e.g.,][]{Madsen2006, Haffner2009, Tomicic2021}. Thus, following the $I_\text{H$\alpha$}$-$I_\text{X}$ correlation, the DIG dominated regions would show lower X-ray emission compared to the dense gas dominated regions.\\

Having established that the SF alone is not able to generate the X-ray emission of JO201, we are left with the other three scenarios presented in §\ref{models}, i.e. stripping of the galactic hot halo, cooling of the ICM onto the galaxy or heating of the ISM through shocks and conduction, mixing of the ICM cooling and ISM heating. The stripping of the hot galactic halo that JO201 may have possessed could produce an extra-planar X-ray emission. However, the face-on orientation of the galaxy makes it difficult to verify this scenario. Nevertheless, many studies \citep[e.g.][]{Larson1980, Bekki2002, Zinger2018,Jaffe2018,Poggianti2019,Gulli2020} highlighted that the majority of galaxies loose their halo in their fall towards the cluster centre, before reaching the virial radius. Given the proximity of JO201 to the cluster centre, well inside the virial radius, we considered that the loss of its halo has already occurred and thus that this scenario is implausible. Therefore, the remaining scenarios are the heating of the ISM (through shocks, conduction or turbulent mixing resulting from the interaction with the ICM), and the cooling of the ICM onto the galaxy (again through conduction or mixing or a combination of the two scenarios). From our spectral analysis we found that the models tested lead to statistically equivalent results (see §\ref{discos}) and for this reason we are not able to discern which of these hypothesis is the better one. Nevertheless, in the following section we sift through each scenario trying to deepen our knowledge about their physics and their implications.

\subsection{Insights from the X-ray spectral models}
In order to get further insights from the X-ray spectral analysis of JO201, we explored in more depth the results obtained from the different spectral models. As first step, we considered the model 1) \texttt{tbabs$\cdot$(apec+apec)} to estimate the mass of the hot galactic plasma. As explained in Section \ref{models}, the first \texttt{apec} represents the ICM emission along the line of sight. For this reason, we can assume that the best-fit values reported in Table \ref{modelli} and related to the galactic component (namely to the second \texttt{apec}) are de-projected quantities. Starting from this consideration, the normalisation of the \texttt{apec} model is linked to the volume of the X-ray emitting plasma through the relation:

\begin{equation}\label{xspecr}
   V=\frac {4 \pi N[D_{\text{A}}(1+z)]^2}{n_{\text{e}}n_{\text{p}}10^{-14}}
\end{equation}

where $N$ is the normalisation parameter of the \texttt{apec} model, $z$ is the redshift, D$_{\text{A}}$=5.26$\cdot$10$^{26}$cm is the angular distance, and $n_{\text{p}}$ is the proton number density, \citep[n$_{\text{p}}$=0.8 n$_{\text{e}}$, e.g.,][]{Gitti2012}. We then followed two parallel procedures. In the first case, we started from the assumption of pressure equilibrium between the ICM (n$_e \sim$10$^{-3}$ cm$^{-3}$ and kT=7.1 $\pm$ 0.2 keV) and the galactic hot (kT $\sim$ 1 keV) plasma, finding that this latter component is characterised by an electron density of n$_e$=(6.4$\pm$0.2)$\cdot$10$^{-3}$cm$^{-3}$. Using the relation \ref{xspecr} we deduced that this plasma is included in a volume, V=(1.8$\pm$0.3)$\cdot$10$^{69}$ cm$^{3}$ and consequently its mass is:

\begin{equation}\label{mgas}
 M_{\text{gas}}= \rho_{\text{gas}}V=1.9 n_{\text{e}} m_{\text{p}}\mu V=(1.1\pm0.2)\cdot 10^{10} M_{\odot}
\end{equation}

where $\rho_{\text{gas}}$ is the mass density of the emitting gas, $\mu=0.6$ is the mean molecular weight and $m_{\text{p}}$ is the proton mass. \\

The second procedure adopted to estimate the mass assumes that the hot galactic plasma occupies a cylindrical volume, with an height equal to the length of the gaseous tail \citep[94 kpc,][]{Bellhouse2019} and an elliptical basis with the same dimensions of the stellar disk \citep[semi-minor axis, $b_{\text{min}}$=$10.7 \pm 0.5$ kpc, and semi-major axis, $b_{\text{max}}$=$13.3 \pm 0.5$ kpc,][]{Gulli2020}, uniformly filled by the hot plasma (i.e. with a filling factor $\phi=1$). As a result of these assumptions, we consider our estimates to be an upper limit of the real volume of the tail. We then estimate the n$_{\text{e}}$ parameter by inverting equation \ref{xspecr}. Also in this case, we can assume that the normalisation of the \texttt{apec} component associated to the galactic emission is a de-projected quantity, because we are modelling the contribution of the ICM along the line of sight with an additional \texttt{apec} component (see §\ref{models}). The electron number density measured in this case is $n_{\text{e}}=(8.0 \pm 0.5)\cdot 10^{-3}$ cm$^{-3}$ and consequently, using equation \ref{mgas}, the mass of the X-ray emitting gas is M$_{\text{gas}}=(9.2 \pm 1.5 )\cdot 10^{9}$ M$_{\odot}$. We note that our estimates are close to the predicted mass of gas with temperature $7\times10^5\text{ K}<T<7\times10^7\text{ K}$ reported by \citet[][]{Tonnesen2011} for an high-velocity jellyfish galaxy ($9.1\times10^{9}$ M$_{\odot}$).  \\

The second model studied in detail is the \texttt{tbabs$\cdot$(apec+mkcflow)}, which describes the ICM cooling scenario. As reported in section §\ref{JO201}, JO201 shows in fact a high fraction of molecular gas and a high SFR, which suggest the presence of an additional source of cold gas in the galaxy. For this reason, we took into consideration the hypothesis that the dominant process responsible for the extended X-ray emission could be the cooling of the ICM onto the galaxy. In this scenario, the increase of the fraction of molecular gas and the corresponding enhancement of the SFR would be fuelled by the continuous accretion of cooling ICM onto the galaxy. In this treatment, we refer to the model 3) (see Table \ref{modelli}) where the metallicity of the \texttt{mkcflow} component is fixed to Z$=$ 0.21 Z$_{\odot}$\footnote{By using the model 4), where the metallicity was let free to vary, both a value of metallicity lower than that of the ICM and a higher error for the mass accretion rate, $\dot{\text{M}}$, were measured.}. The spectral analysis determined a cooling rate of $\dot{M}=1.7^{+0.4}_{-0.6}$ M$_{\odot}$ yr$^{-1}$, which is close to the observed excess of SFR, $\sim3 M_\odot $ yr$^{-1}$, with respect to the median SFR expected for a galaxy of similar mass \citep[][]{Vulcani2018}. This supports the idea that the SFR could be actively fuelled by the cooling gas.\\

Finally, in order to constrain the limits of this pure radiative cooling scenario, we estimated the cooling time of the X-ray emitting plasma with the relation \citep[e.g.,][]{Gitti2012}:

\begin{equation}
    t_{cool}=\frac{5}{2}\frac{kT}{\mu X_H n_e \Lambda(T)}
\end{equation}

where X$_H$=0.71 is the hydrogen mass function, and $\Lambda$(T) is the cooling function (we have interpolated the table by \cite{Southerland1993} as a function of temperature and metallicity). 
By adopting the values of the galactic hot plasma (n$_e$=6.4$\pm$0.2 $\times$ 10$^{-3}$cm$^{-3}$ and kT$\sim$1 keV), the resulting time-scale is t$_{cool} \approx$ 3.6 Gyr, which is not consistent with the time in which the galaxy has experienced the stripping process (0.6-1.2 Gyr, see §\ref{JO201}). This tension could be relieved if the galactic hot plasma has a filamentary structure composed of several clumps with an higher local density (i.e. a volume with $\phi<<1$), which would produce a patchy thermal emission with local peaks where, due to the higher X-ray emissivity, the ICM cools down faster. On the other hand, the discrepancy between the average cooling time estimated above and the stripping time points out that the cooling process could be not only radiative, but it could occur also through conduction or mixing with the ISM \citep[e.g.,][]{Fielding2020}.
\begin{figure*}
\centering
   \includegraphics[scale=0.4]{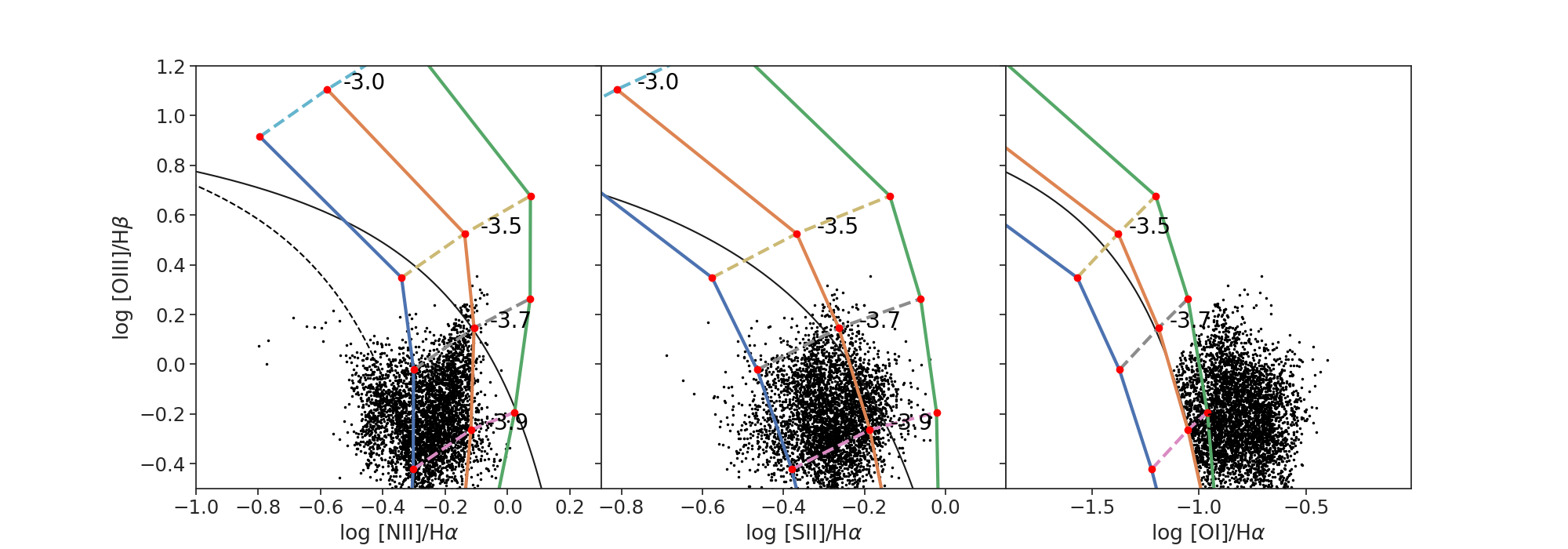}
   \includegraphics[scale=0.4]{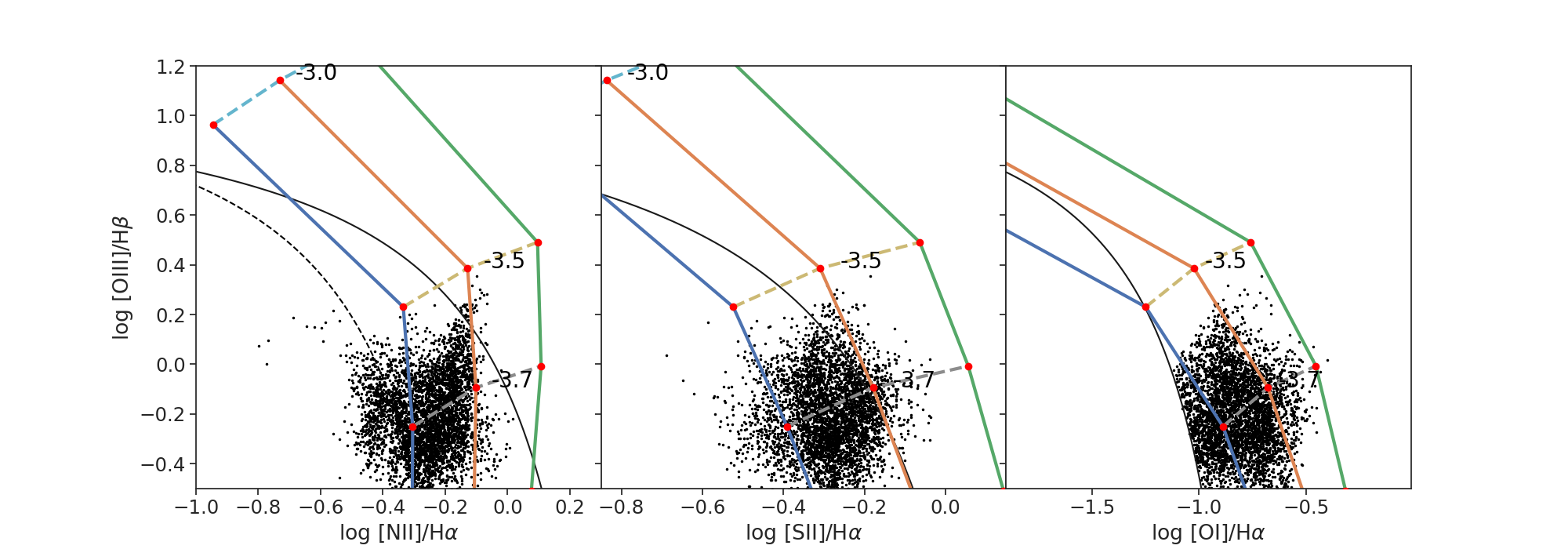}
      \caption{Comparison of line ratios in spaxels classified as LIERs (black dots, distributed as shown in Figure \ref{griglia}) and  CLOUDY photoionization models with different metallicities (from left to right: $Z/Z_\odot = 0.3,0.5,1$) and ionization parameters ($ -4 \le \log U \le -3$), and $n_H$ = 10 cm$^{-3}$. The upper and lower panels shows the results obtained with column densities $\log N_H = 17, 19.5$ respectively. The black solid and dotted curves display the SF/LIER classification by \citet{Kewley06}.}
         \label{cloudy}
   \end{figure*}

\subsection{\texorpdfstring{Origin of [OI]/H$\alpha$ excess}{Origin of [OI]/Halpha excess}}
A number of jellyfish galaxies show an excess of [OI]/H$\alpha$ emission in their tails, whose origin is unclear \citep{Fossati2016, Poggianti2019, Poggianti2019b}. On the basis of our results and those presented in \cite{Poggianti2019}, we speculate that it might be related to the presence of a warm plasma surrounding the stripping filaments, found in §\ref{discos}. To test this hypothesis, we used the CLOUDY C17.02 \citep{Ferland2017} photoionisation code to predict the emission line ratios produced by a cold ISM cloud embedded in a warm plasma, whose properties are constrained by our results (§ \ref{risultati}). To this end, we first derived with  CLOUDY the continuum emitted by a warm plasma described as a collisionally ionised gas with $kT\simeq1$ keV and a hydrogen density $n_H\sim10^{-2}$ cm$^{-3}$: we then computed a grid of models for clouds ionised by this continuum, varying the ionisation parameter ($U$), hydrogen  density ($n_H$) and column density ($N_H$), and metallicities ($Z$).  The ionisation parameter is defined as the ratio of the surface flux of ionizing photons to the hydrogen density, $U=\Phi(H)/(n_H c)$ \citep{Osterbrock2006}. We then looked for the models that provide the best fit to the line ratios measured in  the spaxels classified as LIER.\\

A good agreement is obtained with the following parameters: $n_H \le 10$ cm$^{-3}, \log U \le -3.5$, $Z/Z_\odot \sim 0.3-0.5$, $\log N_H \sim 19.5$. The constraint on the column density is given by the relatively high [OI]/H$\alpha$ ($\sim 0.1$), which requires that  the clouds must be ionisation bounded.
The BPT diagram is displayed in Figure \ref{cloudy} and compares the observed line ratios   with those produced by photoionisation models in a region around the best-fit parameters. \\

We conclude that the ionisation triggered by the warm plasma would be able to reproduce the [OI]/Ha excess in the LIER regions, or at least significantly contribute to it. A similar excess in [OI]/Ha was also found in the stripped gas component of ESO137-001 \citep[][]{Fossati2016} and UGC 6697 \citep[][]{Consolandi2017}, who interpreted it to the presence of shocks. A more detailed analysis, including data from more jellyfish galaxies, will be the subject of a separate paper.

\subsection{Comparison with previous works}
Finally, we compare the results of our work with the previous studies of the galaxies NGC  4569 \citep{Tschoke2001}, NGC 6872 \citep{Machacek2005}, UGC  6697 \citep{Sun2005, Consolandi2017}, ESO 137-001 and ESO110137-002 \citep{Sun2010, Zhang2013} and JW100 \citep{Poggianti2019}. Despite the difference between these galaxies, in terms of stellar mass and velocity, and the properties of the respective environments, the X-ray analysis revealed a number of similarities that can help outlining the origin of the X-ray emitting plasma. Here we list the common aspects:
\begin{itemize}
\item  The X-ray temperature derived for these galaxies with the single temperature model, i.e. the average spectral temperature, is in between 0.7 and 1.0 keV. This value is not consistent with neither the typical temperature of hot gas in normal spirals \citep[kT $\sim0.3$ keV e.g.,][]{Strickland2004,Mineo2012b}, nor the temperature of the surrounding environment;
\item The galaxies for which it was possible to reliably estimate the velocity show similar Mach numbers $\mathcal{M}=2-2.5$; 
\item For the only exception of NGC6872 which is part of a galaxy group, the ICM thermal pressure surrounding these galaxies ranges between 0.8 and 4.2 $\times$ 10$^{-11}$ erg cm$^{-3}$. These values are consistent with threshold of $\sim0.9\times10^{-11}$ erg cm$^{-3}$ defined in \citet{Tonnesen2011} for the formation of bright H$\alpha$ and X-ray filaments.

\end{itemize}
For JW100 and JO201 we identify additional similarities based on the MUSE analysis:
\begin{itemize}
    \item These galaxies show excess of [OI]/H$\alpha$ emission along their tails in correspondence of LIER/composite regions. For JW100 the LIER emission represent the majority of the H$\alpha$ emission outside the stellar disk, whereas in JO201 we observe a lower fraction of LIER emission, but we can not exclude a role of projection effects in this;
   \item We observe in these galaxies a positive spatial correlation between H$\alpha$ and X-ray surface brightness, which suggests that in both galaxies the X-ray emitting plasma follows closely the distribution of the stripped ISM (or vice versa);
    \item Although the different sampling techniques, we observe a similar behaviour between star forming and LIER regions, where the former are generally located to the top-left part of the $I_\text{H$\alpha$}$-$I_\text{X}$ plane with respect to the latter (Figure \ref{ptp}). This could be interpreted either as an excess of H$\alpha$ emission of the star forming region with respect to the LIER regions, or as an excess of X-ray emission of the LIER regions respect to the star forming blobs. However, we note that contrary to \citet{Poggianti2019}, the LIER emission does not seem to correlate with the X-ray. This difference could be due to projection effects, where the orientation of the tail of the galaxy with respect to the line of sight reduces the number of sampling cells at our disposal and strongly affects our results.
\end{itemize}
This comparison pictures a scenario where the X-ray emission arises from the interplay between the ISM and the ICM over the stripped tails (as highlighted by the spatial correlation) that is driven, among the other, by the galactic Mach number and the environmental pressure. This physical framework resembles that of a cold cloud embedded in an hot wind, which has been deeply explored in literature by means of numerical simulations \citep[e.g.,][]{Vollmer2001,Tonnesen2011, Scannapieco2015, Sparre2020, Kanjilal2020}. Albeit all the works predict that the ICM-ISM interplay would lead to the origin of a warm "mixing layer" between the two, the properties and the fate of this are still debated. Therefore, we suggest that the X-ray emitting plasma we studied could correspond to the high-temperature component of the "mixing layer" predicted by the simulations. Furthermore, we argue that the emergence of the [OI]/H$\alpha$ excess in the stripped tails could be another consequence of this ICM-ISM interplay, as consequence of ionisation of the ISM due to the proximity of a warm plasma.  \\

The study of the polarised radio emission of jellyfish galaxies can provide complementary insights into this scenario. \citet[][]{Muller2020} discovered that the magnetic field of another jellyfish galaxy of the GASP sample, JO206, is remarkably ordered along the stripped tail. They suggest that this is consequence of the draped accretion of the ICM onto the galaxy during its motion, which produces a magnetised "draping sheath" lit up by the cosmic rays \citep[][]{Sparre2020}. Transposed to our findings, this ICM draping process could play a role in the origin of the X-ray emitting plasma by favouring the ICM accretion onto the stripped tail of jellyfish galaxies. Exploring in details this scenario requires similar radio studies for JO201 and/or deep X-ray observations of JO206.

\section{Summary and conclusions}
In this paper we have investigated the thermal side of one of the most extreme objects of the GASP sample, namely JO201. This jellyfish galaxy is experiencing a strong stripping process triggered by the ICM of its host cluster A85 and shows an extended X-ray emission. We carried out the first detailed investigation of the X-ray emission of JO201 by means of both a spectral analysis and a point-to point spatial correlation between the H$\alpha$ and the X-ray emission. The dataset used was composed of five Chandra archival observations (total exposure time t$_{exp}\sim$ 187 ks), MUSE H$\alpha$ cubes and emission fraction of the DIG maps. \\

We found that the X-ray luminosity of the galaxy is provided by two contributions. The first one is related to the AGN (L$^{0.5 - 10 \, \text{keV}}_X=2.7 \cdot 10^{41}$erg s$^{-1}$, not corrected for intrinsic absorption), whose emission describes the reflection spectrum of the walls of the torus surrounding the black hole. We reveal an intrinsic photon index of $\Gamma$ =1.7 and a high degree of reflection, as expected in a Compton-Thick AGN.  The second contribution is provided by an extended source associated to a warm plasma (kT$\approx$1 keV) whose X-ray luminosity $L_X^{0.5-10 \, \text{keV}}\approx$ 1.9-4.5 $\cdot$ 10$^{41}$ erg s$^{-1}$ is one order of magnitude higher than the X-ray luminosity expected from the only star formation (L$_X^{0.5-10.0 \, \text{keV}}$ $\sim$ 3.8 $\cdot$ 10$^{40}$ erg s$^{-1}$). The correlation between the H$\alpha$ and X-ray surface brightness emerged from the point to point analysis reveals that this galactic X-ray emitting plasma closely  follows  the  spatial  distribution  of  the ISM. These behaviours suggest that the X-ray extended emission associated to the galaxy results from the local  interplay  between the ICM and the ISM and could be produced by the interface between the two components which envelopes the stripped ISM. \\

In order to investigate the nature of this interplay we tested three different spectral models. However, since they all provide statistically consistent results, we are not able to discern if the X-ray emission is caused by the heating of the ISM, the cooling of the ICM onto the galaxy or the mixing of the ISM and ICM through shocks and conduction. We have therefore deepened our study by taking into consideration the following peculiar properties of JO201 emerged from previous studies: the galaxy shows a high fraction of molecular gas (4-5 times higher than what observed in galaxies of the same mass) and a SFR 0.4–0.5  dex  above  the  main  sequence  of  non-stripped  disk  galaxies;  furthermore, it is characterised by an excess of [OI]/H$\alpha$ emission in correspondence of its tail.\\

Starting from the high fraction of molecular gas and the high SFR, we suggest the presence of an additional source of cold gas which could arise from the cooling of the ICM onto the galaxy. We thus compare the cooling rate obtained from the spectral analysis ($\dot{M}=1.7^{+0.4}_{-0.6}$ M$_{\odot}$ yr$^{-1}$) to the excess of SFR ($\sim3 M_\odot $ yr$^{-1}$), finding that the two values are very close. However, the cooling time of the X-ray emitting plasma, t$_{cool}\approx$3.6 Gyr,  is higher than the stripping time, t$_{stripping}$=1.2-0.6 Gyr. On the one hand this discrepancy could be the result of a a filamentary structure composed of several clumps with an higher local density where, due to the higher X-ray emissivity, the ICM cools down faster. On the other hand, the high cooling time could indicate that the cooling process is not only radiative, but occurs also through conduction or mixing with the ISM. \\

In this work we also investigated the origing of the [OI]/H$\alpha$ excess observed in JO201, as well as in other jellyfish galaxies. Specifically, we tested the possibility that this emission could be originated by the presence of the warm plasma surrounding the stripped ISM. For this reason, we simulated with CLOUDY the behaviour of a cold cloud embedded in a warm plasma, whose thermal properties were constrained by the spectral analysis. From the comparison between the results of the simulations and the line rations observed in JO201 by MUSE, we found that a good agreement is obtained assuming sub-solar values for the metallicity of the warm plasma ($Z=0.3-0.5$). This results is going to be analyzed in more details in a separate paper. \\

The comprehension of the result obtained in this work, as well as their comparison with what revealed from previous studies, is crucial to determine the direction of future analysis. The emerging scenario is that the interaction between the ICM and the ISM originates a warm galactic plasma which is responsible for both the X-ray emission and the [OI]/H$\alpha$ excess in the optical spectrum. Furthermore, the constrain on the low metallicity of the ionising plasma provided by our CLOUDY simulations could indicate that the origin of the warm plasma is strictly connected to the ICM. This suggests that the nature of the interaction between the ICM and the ISM is either the cooling of the ICM or the mixing of these two plasma. On the basis of the results obtained from the radio study of the jellyfish galaxy JO206 \citep[][]{Muller2020}, we speculate that this process could be a direct consequence of the ICM draping and thus that joint radio and X-ray analysis can provide complementary insights of this mechanism. The similarities observed in JW100 and ESO 137-001 (regarding X-ray temperatures, thermal pressure of the surrounding ICM and in the case of JO201 and JW100 also of Mach numbers and excesses of [OI]/H$\alpha$) suggest that the emerging mechanism could be a common feature of jellyfish galaxies. Extending the procedure adopted in this paper and in \cite{Poggianti2019} to other objects of the GASP sample is crucial to confirm this scenario. We also evince that a better overview on the ram pressure stripping process could be reached with both deeper Chandra observations, which provide the opportunity to characterize the thermal properties of jellifish galaxies, and numerical simulations, which, starting from observational constrains could allow to determine the balance between cooling and heating in the ongoing stripping process.

\acknowledgments
We thank the Referee for the useful suggestions that improved the presentation of the work. A.I. thanks C. Stuardi and G. Sabatini for the insightful discussions. Based on observations collected at the European Organization for Astronomical Research in the Southern Hemisphere under ESO programme 196.B-0578. This project has received funding from the European Research Council (ERC) under the Horizon 2020 research and innovation programme (grant agreement N. 833824). 
A.I., B.V., M.G. acknowledge financial contribution from the grant PRIN MIUR 2017 n.20173ML3WW\_001 (PI Cimatti). We acknowledge financial contribution from the INAF main-stream funding programme (PI Vulcani) and from the agreement ASI-INAF n.2017-14-H.0 (P.I. Moretti). YJ acknowledges financial support from CONICYT PAI (Concurso Nacional de Inserci\'on en la Academia 2017) No. 79170132 and FONDECYT Iniciaci\'on 2018 No. 11180558. 
JF acknowledges financial support from the UNAM- DGAPA-PAPIIT IA103520 grant, México. ACCL acknowledges the financial support of the National Agency for Research and Development (ANID) / Scholarship Program / DOCTORADO BECA NACIONAL/2019-21190049

\appendix
\restartappendixnumbering
\section{\texorpdfstring{H$\alpha$ mask}{H alpha mask}}\label{risultha}
\begin{figure}[b!]
    \centering
    \includegraphics[scale=0.4]{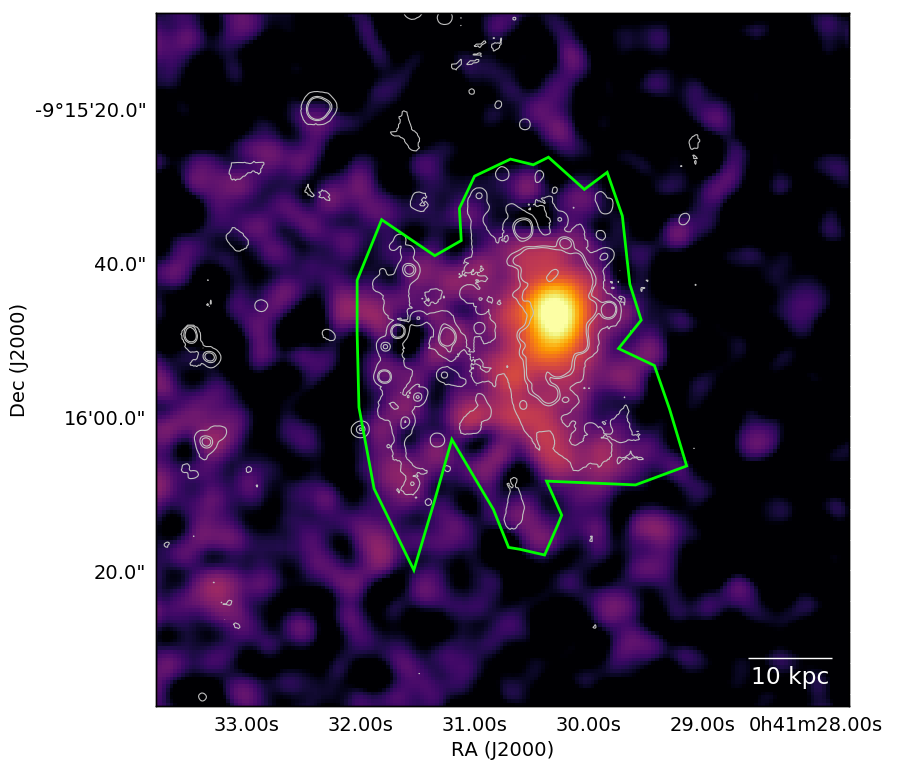}
    \caption{Zoomed-in image of JO201, in green is represented the region used to extract spectra from the H$\alpha$ mask, while in silver is reported the contour of the H$\alpha$ emission \citep[][used as reference]{Bellhouse2019}.}
    \label{mascherinaH}
\end{figure}
We present here the results of the analysis of the X-ray spectrum extracted in a region resembling the H$\alpha$ emission (Figure \ref{mascherinaH}). The procedure followed for the extraction and the fit of the spectra is the same used for the stellar disk mask (see §\ref{discos}). In Table \ref{ha} we report the results of the fits obtained combining the galaxy and the additional components as explained in section §\ref{models}; for two of these models we fixed the T$_{\text{MAX}}$ parameter to the values of the ICM. In particular, the model 5), \texttt{apec+apec}, reveals a temperature of kT=0.79$^{+0.11}_{-0.08}$ keV for the galactic component, which is lower than the ICM temperature. A similar result is also obtained with the model 6), \texttt{apec+cemekl}, where the k$T_{\text{MAX}}$ detected is kT$_{\text{MAX}}$=0.9$^{+0.3}_{-0.3}$ keV, which is lower than the value of the ICM and is compatible with the value found in case 1). For the other two models, 7) \texttt{apec+cemekl} and 8) \texttt{apec+mkcflow}, we set the $T_{\text{MAX}}$ parameter to match the temperature of the ICM. \\

Regarding the metallicity, Z, we note that, when the temperature parameter is free (models 5 and 6), the Z values obtained are lower than that of the ICM. We considered these values implausible, because the X-ray emitting plasma is either the ICM, or the ISM or a mixing of these two components and thus its metallicity is expected to vary between the values of the ICM metallicity (Z$_{\text{ICM}}$=0.21 Z$_{\odot}$, see § \ref{models}) and  of the ISM metallicity (typical values are Z$\sim$1 Z$_{\odot}$). On the contrary, when the temperature is fixed to the value of the ICM (T$_{\text{ICM}}$=7.1 keV, see § \ref{models}) we reveal plausible values of the metallicity, with Z=0.4$^{+0.5}_{-0.2} \, Z_{\odot}$ in model 7 and Z=0.6$\pm 0.2 \, Z_{\odot}$ in model 8. This behaviour attests the presence of a temperature-metallicity degeneracy: the statistic at our disposal does not allow us to reliably estimate at the same time both the temperature and the metallicity parameter of the models used in our analysis.\\

Statistically, it is not possible to discern which emission model is preferred by the fitting procedure because the $\chi^2_{\text{R}}$ values obtained from the four fits are acceptable at 68 \%.
\begingroup
\setlength{\tabcolsep}{2pt}
\begin{table}[h!]\label{tabellaha}

\caption{Fit results for spectra extracted from the H$\alpha$ mask (Figure \ref{mascherinaH}. For this region we resort to the 1B extraction configuration, which models the ICM emission with an additional \texttt{apec} component (see §\ref{models}). The best-fit parameters shown in the second column refer only to the galactic component indicated in the first column (namely 5) \texttt{apec}, 6) and 7) \texttt{cemekl}, 8) \texttt{mkcflow}).} \label{ha}  \

\centering
\begin{tabular}{ccc}

\toprule
                                   Model  & Parameters & \begin{tabular}[c]{@{}c@{}} $\chi^2$/dof \smallskip \\ $\chi^2_{\text{R}}$  \end{tabular}   \\ \toprule
 \begin{tabular}[c]{@{}c@{}}5) \texttt{tbabs$\cdot$(apec+apec)} \end{tabular}& \begin{tabular}[c]{@{}c@{}} kT=0.79$^{+0.11}_{-0.08}$ Z=0.07$^{+0.05}_{-0.03} $ \smallskip \\  F=(3.4$\pm$0.4)$\cdot$10$^{-14}$ L=(2.0 $\pm$ 0.2)$\cdot$10$^{41}$ \end{tabular} &  \begin{tabular}[c]{@{}c@{}}57.02/49 \smallskip \\ 1.164\end{tabular} \\ \cmidrule{1-3} 

\begin{tabular}[c]{@{}l@{}} 6) \texttt{tbabs$\cdot$(apec+cemekl)}\end{tabular} & \begin{tabular}[c]{@{}c@{}}  kT$_{\text{MAX}}$=0.9$^{+0.3}_{-0.3}$ $\alpha$=0.016$^{+3.909}_{-0.017}$\smallskip\\ Z=0.06$^{+0.05}_{-0.02} $ \smallskip\\ F=(2.9$\pm$0.3)$\cdot$10$^{-14}$ L=(1.6$\pm$ 0.2)$\cdot$10$^{41}$ \end{tabular}  & \begin{tabular}[c]{@{}c@{}}54.67/48 \smallskip \\ 1.139\end{tabular} \\ \cmidrule {1-3}  
\begin{tabular}[c]{@{}l@{}}\\ 7) \texttt{tbabs$\cdot$(apec+cemekl)}  \\ kT$_{\text{MAX}}$=7.1 keV (fixed)\end{tabular} & \begin{tabular}[c]{@{}l@{}} $\alpha$=0.01$^{+0.27}_{-0.01}$\smallskip  \\ Z=0.4$^{+0.5}_{-0.2} $  F=(2.7$\pm$0.2)$\cdot$10$^{-14}$ L=(1.3 $\pm$ 0.1)$\cdot$10$^{41}$ \end{tabular}  & \begin{tabular}[c]{@{}c@{}}53.57/49 \smallskip \\ 1.093\end{tabular} \\ \cmidrule{1-3} 
\begin{tabular}[c]{@{}l@{}}8) \texttt{tbabs$\cdot$(apec+mkcflow)}\\ kT$_{\text{MAX}}$=7.1 keV (fixed) \end{tabular}  & \begin{tabular}[c]{@{}l@{}} kT$_{\text{min}}$=0.08$^{+0.17}_{-0.08}$  Z=0.6$^{+0.2}_{-0.2}$  \smallskip \\  $\dot{M}=2.4_{-1.0}^{+3.4}$ \smallskip  \\  F=(6.6$\pm$0.4)$\cdot$10$^{-14}$ L=(3.19 $\pm$ 0.17)$\cdot$10$^{41}$ \end{tabular}  & \begin{tabular}[c]{@{}c@{}}67.78/49 \smallskip \\ 1.383\end{tabular} \\ \bottomrule
\multicolumn{3}{l}{%
  \begin{minipage}{15cm}%
  \medskip  \textbf{Notes:} The unit measure used are: keV for temperature (kT), Z$_{\odot}$ for metallicity (Z), M$_{\odot}$ yr$^{-1}$ for the mass accretion rate ($\dot{\text{M}}$), erg s$^{\text{-1}}$ cm $^{\text{-2}}$ for flux (F) and erg s$^{\text{-1}}$ for luminosity (L). The unabsorbed flux and luminosity refer only to the galactic component and were measured in the 0.5-10.0 keV energy band.
     
  \end{minipage}}
                                \end{tabular}
                                
  \end{table}
\endgroup


\bibliography{sample63}{}
\bibliographystyle{aasjournal}



\end{document}